\documentclass[12pt,onecolumn]{IEEEtran}
\usepackage{setspace}
\usepackage[T1]{fontenc}
\usepackage[latin9]{inputenc}
\usepackage{float}
\usepackage{url}
\usepackage{amsmath}
\usepackage{amsthm}
\usepackage{amssymb}
\usepackage{graphicx}
\usepackage{setspace}
\usepackage{float}
\usepackage{algorithm}
\usepackage{booktabs,subcaption,amsfonts,dcolumn}
\usepackage{amstext}
\usepackage{amsmath}
\usepackage{amsfonts,amssymb}
\usepackage{amsmath,tabu}
\usepackage{amssymb}
\usepackage[dvips]{epsfig}
\usepackage{epsfig}
\usepackage[dvips]{graphicx}
\usepackage{hyperref}
\usepackage{multirow}
\usepackage{multirow} 

\usepackage[labelsep=period]{caption}
\usepackage{subcaption}
\usepackage{float}
\usepackage{ltablex}
\usepackage{cite}
\usepackage{flushend}
\usepackage{algpseudocode}
\usepackage{setspace}

\providecommand{\tabularnewline}{\\}
\floatstyle{ruled}
\newfloat{algorithm}{tbp}{loa}
\providecommand{\algorithmname}{Algorithm}
\floatname{algorithm}{\protect\algorithmname}

\theoremstyle{plain}
\newtheorem{thm}{\protect\theoremname}
\theoremstyle{remark}
\newtheorem{rem}[thm]{\protect\remarkname}

\providecommand{\theoremname}{Theorem}
\usepackage{cite}
\author{Petre Stoica and Prabhu Babu
\thanks{Petre Stoica is with the Division of Systems and Control, Department of Information Technology, Uppsala University, Uppsala, Sweden 75237 and Prabhu Babu is with the Centre for Applied Research in Electronics, Indian Institute of Technology, Delhi 110016, India (email: ps@it.uu.se, Prabhu.Babu@care.iitd.ac.in).
Petre Stoica's work was supported in part by the Swedish Research Council (VR grants 2017-04610,  2016-06079, and 2021-05022).}}

\providecommand{\remarkname}{Remark}

\makeatother

\providecommand{\remarkname}{Remark}
\providecommand{\theoremname}{Theorem}

\begin{document}
\title{Low-rank covariance matrix estimation for factor analysis in anisotropic
noise: application to array processing and portfolio selection}
\maketitle
\begin{abstract}
Factor analysis (FA) or principal component analysis (PCA) models
the covariance matrix of the observed data as $\mathbf{R}=\mathbf{S}\mathbf{S}^{T}+\boldsymbol{\Sigma}$,
where $\mathbf{S}\mathbf{S}^{T}$ is the low-rank covariance matrix
of the factors (aka latent variables) and $\boldsymbol{\Sigma}$ is
the diagonal matrix of the noise. When the noise is anisotropic (aka
nonuniform in the signal processing literature and heteroscedastic
in the statistical literature), the diagonal elements of $\boldsymbol{\Sigma}$
cannot be assumed to be identical and they must be estimated jointly
with the elements of $\mathbf{S}\mathbf{S}^{T}$. The problem of estimating
$\mathbf{S}\mathbf{S}^{T}$ and $\boldsymbol{\Sigma}$ in the above
covariance model is the central theme of the present paper. After
stating this problem in a more formal way, we review the main existing
algorithms for solving it. We then go on to show that these algorithms
have reliability issues (such as lack of convergence or convergence
to infeasible solutions) and therefore they may not be the best possible
choice for practical applications. Next we explain how to modify one
of these algorithms to improve its convergence properties and we also
introduce a new method that we call FAAN (Factor Analysis for Anisotropic
Noise). FAAN is a coordinate descent algorithm that iteratively maximizes
the normal likelihood function, which is easy to implement in a numerically
efficient manner and has excellent convergence properties as illustrated
by the numerical examples presented in the paper. Out of the many
possible applications of FAAN we focus on the following two: direction-of-arrival
(DOA) estimation using array signal processing techniques and portfolio
selection for financial asset management. 
\end{abstract}

\section{Introduction and Problem Statement}

We consider the following data model: 
\begin{equation}
\mathbf{y}_{t}=\mathbf{A}\mathbf{s}_{t}+\mathbf{e}_{t},\;\;\;\;\;t=1,...,N\label{eq:1}
\end{equation}
where \{$\mathbf{y}_{t}\in\mathbb{R}^{n\times1}$\} are the observed
data, $\mathbf{A}\in\mathbb{R}^{n\times r}$ is a transformation matrix,
\{$\mathbf{s}_{t}\in\mathbb{R}^{r\times1}$\} are the (unobserved
or latent) factors, \{$\mathbf{e}_{t}\in\mathbb{R}^{n\times1}$\}
is the noise, and $N$ is the number of observations (or data samples).
We assume that $\left\{ \mathbf{s}_{t}\right\} $ and $\left\{ \mathbf{e}_{t}\right\} $
are uncorrelated to one another and that the covariance matrices of
the two terms in (\ref{eq:1}) are given by: 
\begin{equation}
\begin{aligned}\mathbf{A}\mathbb{E}(\mathbf{s}_{t}\mathbf{s}_{t}^{T})\mathbf{A}^{T} & =\mathbf{A}\mathbf{P}\mathbf{A}^{T}\\
\mathbb{E}(\mathbf{e}_{t}\mathbf{e}_{t}^{T})=\boldsymbol{\Sigma} & =\begin{bmatrix}\sigma_{1}^{2} & 0 & \cdots & 0\\
0 & \sigma_{2}^{2} &  & \vdots\\
\vdots &  & \ddots & 0\\
0 & \cdots & 0 & \sigma_{n}^{2}
\end{bmatrix}
\end{aligned}
\label{eq:2}
\end{equation}
It is also customary to assume that $r<n,$ $\textrm{rank}(\mathbf{A})=\textrm{rank}(\mathbf{P})=r$
and that the data are independent and identically distributed. Under
these assumptions the covariance matrix of $\mathbf{y}_{t}$ can be
written as: 
\begin{equation}
\mathbf{R}=\mathbf{S}\mathbf{S}^{T}+\boldsymbol{\Sigma},\;\mathbf{S}\in\mathbb{R}^{n\times r},\;\textrm{rank}(\mathbf{S})=r\label{eq:3}
\end{equation}
(where $\mathbf{S}$ depends on $\mathbf{A}$ and $\mathbf{P}$ in
an obvious way).

The principal problem dealt with in this paper is the estimation of
$\mathbf{S}\mathbf{S}^{T}$ and $\boldsymbol{\Sigma}$ in the above
covariance model from data $\{\mathbf{y}_{t}\}_{t=1}^{N}$. Note that
an orthogonal linear transformation of $\mathbf{S}^{T}$ in (\ref{eq:1})
does not change $\mathbf{R}$ (indeed $\mathbf{S}\mathbf{S}^{T}=\mathbf{S}\mathbf{Q}\mathbf{Q}^{T}\mathbf{S}^{T}$
for any $\mathbf{Q}$ that satisfies $\mathbf{Q}\mathbf{Q}^{T}=\mathbf{I}$)
and therefore only the range of $\mathbf{S},\;\mathcal{R}(\mathbf{S})$,
can be uniquely estimated by means of a covariance fitting procedure.
This is not a problem for most applications as in general an estimate
of $\mathcal{R}(\textbf{S})$ or $\mathbf{S}\mathbf{S}^{T}$ suffices.
However, as a preamble to the discussion in Section \ref{sec:3},
we note here that the model in (\ref{eq:1}) has other possible non-trivial
ambiguities in the sense that even $\mathbf{S}\mathbf{S}^{T}$ and
$\boldsymbol{\Sigma}$ corresponding to a given covariance matrix
might not be unique. This is an important aspect of the parametrization
in (\ref{eq:1}), from both a theoretical and practical standpoint,
which will be discussed in Section \ref{sec:3}.

The central problem of FA or PCA is precisely the covariance matrix
estimation problem stated above and there is a huge literature about
it that goes back to the beginning of the previous century (see, e.g.,
\cite{spearman,thurstone,frisch,lederman,anderson}). Early applications
were in psychometrics, econometrics and, more general, statistics;
while more recent ones are in signal processing, marketing, biological
sciences and, more general, big data analysis and machine learning.
Even a partial review of the practical and theoretical works on FA
and PCA published along the years is an immense task that we will
not undertake. Instead we will refer only to the papers that we found
to be directly relevant to the discussion and the approach presented
in this paper. These papers, which include \cite{Tryphon,ciccone,liao,HPCA,bekker,shapiro,pesavento}
and are discussed in the sections where they are most relevant, contain
many references that can be consulted by the reader who is interested
in the historical and more recent developments of the basic ideas
and methods of FA and PCA.

\section{\label{sec:2}Critical review of previous methods}

This section discusses three recently proposed methods for estimating
the parameters of the covariance model in (\ref{eq:3}). While these
methods appear to be state-of-the-art, we show that they are not without
problems.

\subsection{Frobenius Norm Method}

We will use the acronym $\textrm{FNM}_{\textrm{o}}$ to designate
this method, with the index ``o'' being used to indicate the original
version and distinguish it from the modified version presented later
in the paper, which will be simply called FNM. Let 
\begin{equation}
\hat{\mathbf{R}}=\frac{1}{N}\sum_{t=1}^{N}\mathbf{y}_{t}\mathbf{y}_{t}^{T}
\end{equation}
denote the sample covariance matrix (SCM). Quite possibly the most
direct method for estimating $\mathbf{S}\mathbf{S}^{T}$ and $\boldsymbol{\Sigma}$
is via solving the following minimization problem: 
\begin{equation}
\begin{array}{ll}
\underset{\mathbf{S}\mathbf{S}^{T},\boldsymbol{\Sigma}}{\min}\; & \lVert\hat{\mathbf{R}}-\mathbf{S}\mathbf{S}^{T}-\boldsymbol{\Sigma}\rVert\end{array}\label{eq:5}
\end{equation}
where $\lVert\cdot\rVert$ denotes the Frobenius norm. The following
coordinate descent algorithm can be used to find the minimizers of
the function $g\triangleq\lVert\hat{\mathbf{R}}-\mathbf{S}\mathbf{S}^{T}-\boldsymbol{\Sigma}\rVert$
in (\ref{eq:5}).

\begin{algorithm}[H]
\caption{$\textrm{FNM}_{\textrm{o}}$ algorithm}

\textbf{Input:} $\hat{\mathbf{R}}$, $r$ and $\hat{\boldsymbol{\Sigma}}_{0}$.

For $i=0,1,2,\ldots,$ do: 
\begin{itemize}
\item Let $e_{1}\geq e_{2}\geq e_{3}\geq\cdots\geq e_{n}$ denote the eigenvalues
of ($\hat{\mathbf{R}}-\hat{\boldsymbol{\Sigma}}_{i}$) and let $\mathbf{w}_{1},\mathbf{w}_{2},...,\mathbf{w}_{n}$
be the corresponding eigenvectors. Then, by a well-known result, the
minimizer $\widehat{\mathbf{S}\mathbf{S}^{T}}$ of $g$, for given
$\boldsymbol{\Sigma}=\hat{\boldsymbol{\Sigma}}_{i}$, is 
\begin{equation}
\widehat{\mathbf{S}\mathbf{S}^{T}}_{i}=\begin{bmatrix}\mathbf{w}_{1} & \cdots & \mathbf{w}_{r}\end{bmatrix}\begin{bmatrix}e_{1} & 0 & \cdots & 0\\
0 & e_{2} &  & \vdots\\
\vdots &  & \ddots & 0\\
0 & \cdots & 0 & e_{r}
\end{bmatrix}\begin{bmatrix}\mathbf{w}_{1}^{T}\\
\vdots\\
\mathbf{w}_{r}^{T}
\end{bmatrix}\label{eq:6}
\end{equation}
\item For given $\mathbf{S}\mathbf{S}^{T}=\widehat{\mathbf{S}\mathbf{S}^{T}}_{i}$,
the minimizer $\hat{\boldsymbol{\Sigma}}$ of $g$ is 
\begin{equation}
\hat{\boldsymbol{\Sigma}}_{i+1}=\textrm{diag}(\hat{\mathbf{R}}-\widehat{\mathbf{S}\mathbf{S}^{T}}_{i}),\label{eq:7}
\end{equation}
where, for any square matrix $\mathbf{M}\in\mathbb{R}^{n\times n}$,
$\textrm{diag}(\mathbf{M})=\footnotesize\begin{bmatrix}M_{11} & 0 & \cdots & 0\\
0 & M_{22} &  & \vdots\\
\vdots &  & \ddots & 0\\
0 & \cdots & 0 & M_{nn}
\end{bmatrix}.$ 
\item If $\left(g_{i-1}-g_{i}\right)\slash g_{i}\leq\epsilon$ (where $g_{-1}=\infty$
and $\epsilon=10^{-3}$ for example) then goto Output; else $i\rightarrow i+1$
and iterate. 
\end{itemize}
\textbf{Output}: $\widehat{\mathbf{S}\mathbf{S}^{T}}$ and $\widehat{\boldsymbol{\Sigma}}$ 
\end{algorithm}

The above simple algorithm certainly has been known for quite some
time to researchers working in the field. It was rediscovered many
times, for example, relatively recently in \cite{liao} where an unnecessarily
long derivation of it was presented. In an even more recent paper,
\cite{HPCA}, an algorithm was proposed that coincides with the above
one with the initialization $\hat{\boldsymbol{\Sigma}}_{0}=\textrm{diag}(\hat{\mathbf{R}})$.
As we explain next, in the latter algorithm (see below) the step of
updating $\boldsymbol{\Sigma}$ was implicit rather than explicit,
which made it somewhat difficult to see that this algorithm was a
variant of $\textrm{FNM}_{\textrm{o}}$.

\begin{algorithm}[H]
\caption{A variant of the $\textrm{FNM}_{\textrm{o}}$ algorithm}

\textbf{Input:} $\hat{\mathbf{R}}$, $r$ and $\widehat{\mathbf{S}\mathbf{S}^{T}}_{0}=\boldsymbol{0}$.

For $i=0,1,2,\ldots$, do: 
\begin{itemize}
\item Let $\left\{ e_{k},\mathbf{w}_{k}\right\} $ be as defined in $\textrm{FNM}_{\textrm{o}}$
but for the matrix 
\[
\textrm{diag}_{0}\left(\hat{\mathbf{R}}\right)+\textrm{diag}\left(\widehat{\mathbf{S}\mathbf{S}^{T}}_{i}\right)
\]
where the operator $\textrm{diag}_{0}\left(\cdot\right)$ replaces
the diagonal of the argument with zeros. 
\item Update $\widehat{\mathbf{S}\mathbf{S}^{T}}_{i+1}$ as in (\ref{eq:6})
and iterate until a convergence criterion is satisfied. 
\end{itemize}
\textbf{Output}: $\widehat{\mathbf{S}\mathbf{S}^{T}}$ 
\end{algorithm}

Because (see \eqref{eq:7}) 
\begin{equation}
\hat{\mathbf{R}}-\hat{\boldsymbol{\Sigma}}_{i+1}=\hat{\mathbf{R}}-\textrm{diag}(\hat{\mathbf{R}}-\widehat{\mathbf{S}\mathbf{S}^{T}}_{i})=\textrm{diag}_{0}(\hat{\mathbf{R}})+\textrm{diag}(\widehat{\mathbf{S}\mathbf{S}^{T}}_{i}),
\end{equation}
it readily follows that the algorithm of \cite{HPCA} is identical
to the $\textrm{FNM}_{o}$ with $\hat{\boldsymbol{\Sigma}}_{o}=\textrm{diag}(\hat{\mathbf{R}})$
(a fact that was apparently overlooked in \cite{HPCA}).

Our tests of $\textrm{FNM}_{\textrm{o}}$ have shown that quite often
this algorithm converges to an infeasible solution, i.e., one for
which either $\widehat{\mathbf{S}\mathbf{S}^{T}}$ or $\hat{\boldsymbol{\Sigma}}$
are indefinite matrices. To illustrate this issue of $\textrm{FNM}_{\textrm{o}}$,
we will compare it in what follows with a slightly modified version
that we call $\textrm{FNM}$ for simplicity. In the cases that were
problematic for $\textrm{FNM}_{\textrm{o}}$ we have observed that
the matrix $\textrm{diag}(\hat{\mathbf{R}}-\widehat{\mathbf{S}\mathbf{S}^{T}}_{i})$
had negative elements. Consequently, we replace (\ref{eq:7}) in $\textrm{FNM}_{\textrm{o}}$
by 
\begin{equation}
\hat{\boldsymbol{\Sigma}}_{i+1}=\textrm{diag}(\hat{\mathbf{R}}-\widehat{\mathbf{S}\mathbf{S}^{T}}_{i})_{+}
\end{equation}
where the operator replaces $(\cdot)_{+}$ replaces the negative elements
on the diagonal of the argument with zeros and leaves the positive
elements unchanged.

\emph{Example 1: Illustration of the infeasibility issue of $\textrm{FNM}_{\textrm{o}}$ }

We apply $\textrm{FNM}_{\textrm{o}}$ and $\textrm{FNM}$ to the following
randomly generated sample covariance matrix ($n=6)$: 
\begin{equation}
\footnotesize\hat{\mathbf{R}}=\begin{bmatrix}1.0973 & -0.2093 & 0.9481 & -1.4471 & 1.7815 & -0.7927\\
-0.2093 & 4.4978 & 0.4230 & 4.4947 & -1.7959 & 3.2707\\
0.9481 & 0.4230 & 3.5566 & 0.1260 & 0.5104 & -2.3557\\
-1.4471 & 4.4947 & 0.1260 & 7.5986 & -3.0046 & 1.4273\\
1.7815 & -1.7959 & 0.5104 & -3.0046 & 6.8526 & -2.9834\\
-0.7927 & 3.2707 & -2.3557 & 1.4273 & -2.9834 & 7.9070
\end{bmatrix}
\end{equation}
for $r=2$ and the following two initializations (the first one was
suggested in \cite{liao}, the second one was used in \cite{HPCA}):
\begin{equation}
\begin{aligned}\hat{\boldsymbol{\Sigma}}_{0} & =\mathbf{I}\\
\hat{\boldsymbol{\Sigma}}_{0} & =\textrm{diag}(\hat{\mathbf{R}})
\end{aligned}
\end{equation}
The two algorithms provide the following estimates:\\
 \textbf{$\textrm{\textbf{FNM}}_{\textrm{o}}$} $(\hat{\boldsymbol{\Sigma}}_{0}=\mathbf{I})$
\begin{equation}
\footnotesize\begin{aligned}\widehat{\mathbf{S}\mathbf{S}^{T}}=\begin{bmatrix}0.2804 & -0.8385 & 0.0785 & -1.4787 & 0.6368 & -1.0128\\
-0.8385 & 2.5086 & -0.2498 & 4.3938 & -1.9102 & 3.1198\\
0.0785 & -0.2498 & 0.3621 & 0.2351 & 0.3035 & -2.3288\\
-1.4787 & 4.3938 & 0.2351 & 9.0372 & -3.1198 & 1.4390\\
0.6368 & -1.9102 & 0.3035 & -3.1198 & 1.4926 & -3.0535\\
-1.0128 & 3.1198 & -2.3288 & 1.4390 & -3.0535 & 15.9575
\end{bmatrix},\;\hat{\boldsymbol{\Sigma}}=\begin{bmatrix}0.8169 & 0 & 0 & 0 & 0 & 0\\
0 & 1.9891 & 0 & 0 & 0 & 0\\
0 & 0 & 3.1945 & 0 & 0 & 0\\
0 & 0 & 0 & -1.4386 & 0 & 0\\
0 & 0 & 0 & 0 & 5.3600 & 0\\
0 & 0 & 0 & 0 & 0 & -8.0505
\end{bmatrix}\end{aligned}
\label{eq:12-1}
\end{equation}
$\textrm{\textbf{FNM}}(\hat{\boldsymbol{\Sigma}}_{o}=\mathbf{I})$
\begin{equation}
\footnotesize\begin{aligned}\widehat{\mathbf{S}\mathbf{S}^{T}}=\begin{bmatrix}0.3202 & -0.9520 & 0.1943 & -1.3001 & 0.7656 & -1.1482\\
-0.9520 & 2.9223 & -0.3419 & 4.3355 & -2.2416 & 2.8172\\
0.1943 & -0.3419 & 0.7264 & 0.4222 & 0.5551 & -2.2374\\
-1.3001 & 4.3355 & 0.4222 & 7.6905 & -2.9293 & 1.5966\\
0.7656 & -2.2416 & 0.5551 & -2.9293 & 1.8444 & -2.9748\\
-1.1482 & 2.8172 & -2.2374 & 1.5966 & -2.9748 & 8.0179
\end{bmatrix},\;\hat{\boldsymbol{\Sigma}}=\begin{bmatrix}0.7771 & 0 & 0 & 0 & 0 & 0\\
0 & 1.5755 & 0 & 0 & 0 & 0\\
0 & 0 & 2.8302 & 0 & 0 & 0\\
0 & 0 & 0 & 0 & 0 & 0\\
0 & 0 & 0 & 0 & 5.0082 & 0\\
0 & 0 & 0 & 0 & 0 & 0
\end{bmatrix}\end{aligned}
\label{eq:13}
\end{equation}
\textbf{$\textrm{FNM}_{\textrm{o}}$} ($\hat{\boldsymbol{\Sigma}}_{0}=\textrm{diag}(\hat{\mathbf{R}})$)
\begin{equation}
\footnotesize\begin{aligned}\widehat{\mathbf{S}\mathbf{S}^{T}}=\begin{bmatrix}0.2803 & -0.8384 & 0.0793 & -1.4782 & 0.6370 & -1.0141\\
-0.8384 & 2.5084 & -0.2514 & 4.3942 & -1.9107 & 3.1185\\
0.0793 & -0.2514 & 0.3643 & 0.2365 & 0.3056 & -2.3282\\
-1.4782 & 4.3942 & 0.2365 & 9.0485 & -3.1193 & 1.4393\\
0.6370 & -1.9107 & 0.3056 & -3.1193 & 1.4938 & -3.0537\\
-1.0141 & 3.1185 & -2.3282 & 1.4393 & -3.0537 & 15.8590
\end{bmatrix},\;\hat{\boldsymbol{\Sigma}}=\begin{bmatrix}0.8170 & 0 & 0 & 0 & 0 & 0\\
0 & 1.9893 & 0 & 0 & 0 & 0\\
0 & 0 & 3.1924 & 0 & 0 & 0\\
0 & 0 & 0 & -1.4499 & 0 & 0\\
0 & 0 & 0 & 0 & 5.3588 & 0\\
0 & 0 & 0 & 0 & 0 & -7.9520
\end{bmatrix}\end{aligned}
\label{eq:14-1}
\end{equation}
\textbf{$\textrm{FNM}$} ($\hat{\boldsymbol{\Sigma}}_{0}=\textrm{diag}(\hat{\mathbf{R}}$))
\begin{equation}
\footnotesize\begin{aligned}\widehat{\mathbf{S}\mathbf{S}^{T}}=\begin{bmatrix}0.3202 & -0.9520 & 0.1943 & -1.3001 & 0.7656 & -1.1482\\
-0.9520 & 2.9223 & -0.3419 & 4.3355 & -2.2416 & 2.8172\\
0.1943 & -0.3419 & 0.7264 & 0.4222 & 0.5551 & -2.2374\\
-1.3001 & 4.3355 & 0.4222 & 7.6905 & -2.9293 & 1.5966\\
0.7656 & -2.2416 & 0.5551 & -2.9293 & 1.8444 & -2.9748\\
-1.1482 & 2.8172 & -2.2374 & 1.5966 & -2.9748 & 8.0179
\end{bmatrix},\;\hat{\boldsymbol{\Sigma}}=\begin{bmatrix}0.7771 & 0 & 0 & 0 & 0 & 0\\
0 & 1.5755 & 0 & 0 & 0 & 0\\
0 & 0 & 2.8302 & 0 & 0 & 0\\
0 & 0 & 0 & 0 & 0 & 0\\
0 & 0 & 0 & 0 & 5.0082 & 0\\
0 & 0 & 0 & 0 & 0 & 0
\end{bmatrix}\end{aligned}
\label{eq:15-1}
\end{equation}
In (\ref{eq:12-1}) and (\ref{eq:14-1}) $\hat{\boldsymbol{\Sigma}}$
is indefinite. In contrast to this, both $\hat{\boldsymbol{\Sigma}}$
and $\widehat{\mathbf{S}\mathbf{S}^{T}}$ are positive semidefinite
in (\ref{eq:13}) and (\ref{eq:15-1}), as required, and this despite
the fact that FNM constrains only $\boldsymbol{\Sigma}$ to have nonnegative
elements. We have not encountered a case in which the matrix $\widehat{\mathbf{S}\mathbf{S}^{T}}$
obtained with FNM was indefinite, but if this happens then FNM can
be run with the constraint $\widehat{\mathbf{S}\mathbf{S}^{T}}\succeq0$
as well (however doing so may decrease the rank under $r$). Also
note that FNM has converged to exactly the same matrices $\widehat{\textbf{S}\textbf{S}^{T}}$
and $\hat{\mathbf{\Sigma}}$ from both initial points, see \eqref{eq:13}
and \eqref{eq:15-1}, whereas the $\text{FNM}_{{o}}$ estimates in
\eqref{eq:12-1} and \eqref{eq:14-1} are somewhat different (possibly
due to a slower convergence rate).

\subsection{Maximum likelihood (ML) method for deterministic $\left\{ \mathbf{s}_{t}\right\} $}

Under the assumption that the latent variables $\left\{ \mathbf{s}_{t}\right\} $
are deterministic and the noise $\left\{ \mathbf{e}_{t}\right\} $
in (\ref{eq:1}) has a normal distribution, the negative log-likelihood
function of $\left\{ \mathbf{y}_{t}\right\} $ is given by (to within
a constant): 
\begin{equation}
N\sum_{k=1}^{n}\ln\sigma_{k}^{2}+\sum_{t=1}^{N}\sum_{k=1}^{n}\frac{1}{\sigma_{k}^{2}}(y_{tk}-\mathbf{a}_{k}^{T}\mathbf{s}_{t})^{2}\label{eq:12}
\end{equation}
where $y_{tk}$ is the $k$th element of $\mathbf{y}_{t}$ and $\mathbf{a}_{k}^{T}$
is the $k$th row of $\mathbf{A}$. The so-called nonuniform ML method,
proposed in $[12]$ and reviewed in \cite{liao}, tries to minimize
(\ref{eq:12}) to estimate both $\left\{ \sigma_{k}^{2}\right\} $
and $\left\{ \mathbf{s}_{t}\right\} $ as well as $\mathbf{A}$ (in
a parameterized form, see the cited references). The problem is that
the ML estimate of the aforementioned parameters does not exist, which
can be seen by an analysis similar to the one in \cite{stoica1}.
Indeed, let us choose $\left\{ \mathbf{s}_{t}\right\} $ so that 
\begin{equation}
y_{t1}=\mathbf{a}_{1}^{T}\mathbf{s}_{t}\;\;\;\;t=1,2,\ldots,N
\end{equation}
Then the summand that depends on $\sigma_{1}^{2}$ in the second term
of (\ref{eq:12}) is equal to zero and consequently if we let $\sigma_{1}^{2}\rightarrow0$,
the function in (\ref{eq:12}) tends to $-\infty$. This simple observation,
showing that the function in (\ref{eq:12}) is unbounded from below,
implies that there is no global minimum and therefore the ML estimate
does not exist (which was intuitively expected given the excessive
number of unknowns in (\ref{eq:12})).

\subsection{ML method for stochastic $\left\{ \mathbf{s}_{t}\right\} $}

Under the assumption that both $\left\{ \mathbf{s}_{t}\right\} $
and $\left\{ \mathbf{e}_{t}\right\} $ have normal distributions with
means zero and covariances $\mathbf{P}$ and \textbf{$\boldsymbol{\Sigma}$,}
the negative log-likelihood function of $\left\{ \mathrm{\mathbf{y}}_{t}\right\} $
is given by (neglecting some multiplicative and additive constants):
\begin{equation}
\textrm{Tr}\left(\hat{\mathbf{R}}\mathbf{R}^{-1}\right)+\ln\left|\mathbf{R}\right|\label{eq:14}
\end{equation}
where $\mathbf{R}$ is as defined in (\ref{eq:3}). The statistical
properties of the maximum likelihood estimates that minimize (\ref{eq:14})
have been studied, for example, in \cite{bai} where the consistency,
convergence rate, limiting distribution and asymptotic efficiency
were proved. 

An algorithm for minimizing the above function was proposed in \cite{liao} where it was called Iterative ML Subspace Estimation (IMLSE). The
IMLSE minimizes (\ref{eq:14}) with respect to (wrt) $\mathbf{S}\mathbf{S}^{T}$,
for given $\boldsymbol{\Sigma}$, and then updates $\boldsymbol{\Sigma}$,
for given $\mathbf{S}\mathbf{S}^{T}$, using a fixed-point iteration.
This combination of a partial minimizer (more exactly, a stationary
point) and a fixed-point iterative scheme is not guaranteed to monotonically
decrease the loss and in fact it may fail to converge as shown in
the Example 2 below. 

Before presenting the example we note that the IMLSE method suggested
in \cite{liao} is not new. Indeed, this type of method, which basically
tries to solve the equations satisfied by the stationary points of
the likelihood function by means of a fixed-point iteration, has been
known for many years and its problems are well documented in the literature
(see e.g., \cite{joreskog,lawley} and references therein). Our experience is
that the method works in most cases if $n$ is small and $r$ is considerably
smaller than $n$. As $n$ and $r$ increase this method fails to
work properly more and more frequently. The main problem appears to
be that, similarly to $\textrm{FNM}_{\textrm{o}}$, the iterates can
step out the feasibility set and when this happens a matrix square-root
computed in the algorithm gets imaginary elements. Setting those problematic
elements to zero sometimes fixes the problem but not always (for instance,
it did not eliminate the problem in the case illustrated in Fig.
1). For a more detailed critical discussion on this type of method
for minimizing the likelihood function in (\ref{eq:14}) see \cite{joreskog}.\\
\emph{Example 2: Illustration of the convergence problem of IMLSE }

\begin{figure}[ht]
\begin{centering}
\begin{tabular}{c}
\includegraphics[width=8cm,height=6cm]{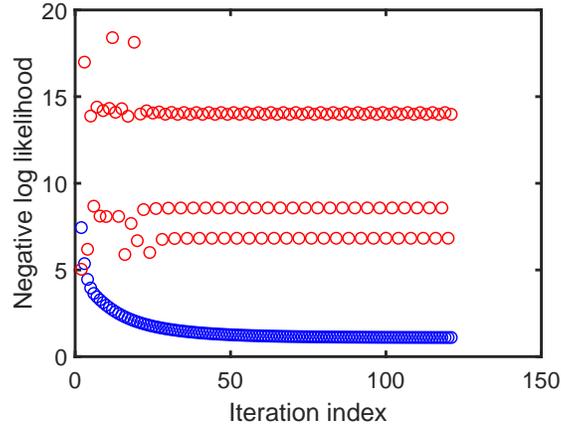}\tabularnewline
\tabularnewline
\end{tabular}
\par\end{centering}
\caption{\label{fig:1} The cost function in (\ref{eq:14}) vs the iteration
number for FAAN (blue circles) and IMLSE (red circles)}
\end{figure}

We run IMLSE and FAAN (which will be introduced later on) for the
sample covariance matrix in (\ref{eq:R}) with $r=3$ and initial
value $\hat{\boldsymbol{\Sigma}}_{0}=\mathbf{I}$. Fig. \ref{fig:1}
shows the variation of the loss in (\ref{eq:14}) versus the iteration
number for both the algorithms. One can see that the loss corresponding
to the IMLSE initially varies quite a bit and then, as the number
of iterations increases, it oscillates between three values. In stark
contrast to this, FAAN monotonically decreases the loss and has no
convergence problem at all.

\begin{equation}
\footnotesize\begin{aligned}\hat{\mathbf{R}}=\begin{bmatrix}5.9022 & 3.2245 & 7.3856 & 4.7320 & 4.7804\\
3.2245 & 2.1207 & 3.9317 & 2.5892 & 1.6077\\
7.3856 & 3.9317 & 9.3943 & 5.9126 & 5.6763\\
4.7320 & 2.5892 & 5.9126 & 3.9139 & 3.6792\\
4.7804 & 1.6077 & 5.6763 & 3.6792 & 10.4673
\end{bmatrix}\end{aligned}
\label{eq:R}
\end{equation}

The FAAN algorithm is derived and discussed in Section IV. Before
presenting FAAN, however, we need to discuss the uniqueness of the
parametrization in (\ref{eq:3}). 

\section{\label{sec:3}Identifiability of the parametrization}

The parametrization of $\mathbf{R}$ in (\ref{eq:3}) is said to be
globally identifiable if it is unique. In other words, if $\mathbf{R}=\mathbf{S}\mathbf{S}^{T}+\boldsymbol{\Sigma}$,
then there is no other $\tilde{\boldsymbol{\Sigma}}\neq\boldsymbol{\Sigma}$
(and $\widetilde{\mathbf{S}\mathbf{S}^{T}}\neq\mathbf{S}\mathbf{S}^{T}$,
with $\textrm{rank}(\widetilde{\mathbf{S}\mathbf{S}^{T}})\leq r$)
such that $\mathbf{R}=\widetilde{\mathbf{S}\mathbf{S}^{T}}+\tilde{\boldsymbol{\Sigma}}$.
If this is true only in a small neighbourhood of $\mathbf{\Sigma}$,
then the parametrization is said to be locally identifiable.

Identifiability (or uniqueness) of the parametrization is an important
property without which the estimate of $\mathbf{S}\mathbf{S}^{T}$
and $\boldsymbol{\Sigma}$ (for example, obtained by minimizing the
loss in (\ref{eq:14})) may be unreliable. More specifically, if our
goal is to obtain a more accurate estimate of the covariance matrix
than $\widehat{\mathbf{R}}$, then a possible lack of identifiability
may be acceptable, as all estimates of $\mathbf{S}\mathbf{S}^{T}$
and $\boldsymbol{\Sigma}$ yield the same estimate of $\mathbf{R}$.
However, for applications in which we need an estimate of $\mathcal{R}\left(\mathbf{S}\right)$
(such as FA and array signal processing, see Section \ref{sec:5}),
identifiability is an essential property as without it the estimate
of $\mathcal{R}\left(\mathbf{S}\right)$ may be heavily biased.

The identifiability of any parametrization is related to the interplay
between the number of free parameters (or unknowns) of the model,
let us say $n_{m}$, and the number of constraints imposed on them,
denoted $n_{c}$. In the present case, 
\begin{equation}
n_{c}=\frac{n(n+1)}{2}\label{eq:15}
\end{equation}
and $n_{m}$ can be shown to be 
\begin{equation}
n_{m}=(n-r)r+\frac{r(r+1)}{2}+n\label{eq:16}
\end{equation}
(to verify (\ref{eq:16}), use an LQ decomposition of an $r\times r$
invertible block of $\mathbf{S}$; there remain $\frac{r(r+1)}{2}$
free parameters in the said block, and the rest of the matrix $\mathbf{S}$
has $(n-r)r$ free parameters; furthermore this parametrization of
$\mathbf{S}$ is unique; and the last term in (\ref{eq:16}) is due
to $\boldsymbol{\Sigma}$). From (\ref{eq:15}) and (\ref{eq:16})
we have that: 
\begin{equation}
2(n_{c}-n_{m})=r^{2}-r(2n+1)+(n^{2}-n)\label{eq:17}
\end{equation}
The roots of the quadratic polynomial in (\ref{eq:17}) are 
\begin{equation}
r_{1,2}=\frac{2n+1\pm\sqrt{8n+1}}{2}
\end{equation}
Only one of $r_{1,2}$ is less than $n$ (as it should be): 
\begin{equation}
r_{L}=\frac{2n+1-\sqrt{8n+1}}{2}\label{eq:19}
\end{equation}
Note that for $r=0$, the polynomial in (\ref{eq:17}) takes on a
positive value, viz. $n(n-1)$. It follows that: 
\begin{equation}
r\leq r_{L}\implies n_{c}\geq n_{m}
\end{equation}
and 
\begin{equation}
r>r_{L}\implies n_{c}<n_{m}
\end{equation}
Therefore, it is intuitively expected that the parametrization is
identifiable for $r\leq r_{L}$ and unidentifiable for $r>r_{L}$.
The index $L$ of $r_{L}$ is motivated by the fact that the author
of \cite{lederman} was the first to introduce the bound in (\ref{eq:19})
as well as the previous conjecture about identifiability based on
whether $r$ is smaller or larger than $r_{L}$.\\
 The above conjecture was shown to be true in \cite{bekker} where
the following result was proved (below ``generically'' means for
almost any instance of $\mathbf{S}\mathbf{S}^{T}$ and $\boldsymbol{\Sigma}$,
except for some matrices $\mathbf{S}\mathbf{S}^{T}$ and $\boldsymbol{\Sigma}$
that lie in a set of measure zero).\\
 \textbf{Identifiability result} 
\begin{enumerate}
\item[a)] If $r<r_{L}$, then the model (\ref{eq:3}) is generically globally
identifiable. 
\item[b)] If $r>r_{L}$, then the model is generically locally (and hence globally)
unidentifiable. 
\item[c)] If $r=r_{L}$, the model is generically locally identifiable. 
\end{enumerate}
It can be seen from (\ref{eq:19}) that $r_{L}$ approaches $n$,
as $n$ increases. However for small $n$, $r_{L}$ is significantly
smaller than $n$, see Fig. 2. 
\begin{figure}[ht]
\begin{centering}
\begin{tabular}{c}
\includegraphics[width=8cm,height=6cm]{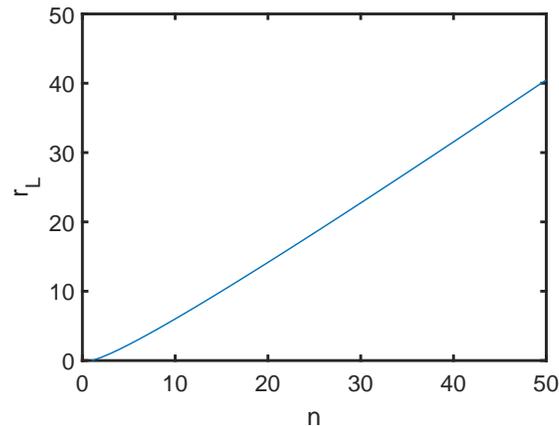} \tabularnewline
\end{tabular}
\par\end{centering}
\caption{The bound $r_{L}$ vs $n$. }
\end{figure}

An interesting question in the present context, although not strictly
related to identifiablity, is whether any $\widehat{\mathbf{R}}$
can be written in the form (\ref{eq:3}) for a given $r$. In other
words, given $\widehat{\mathbf{R}}$ and $r$ can we find $\mathbf{S}$
and $\boldsymbol{\Sigma}$ such that $\widehat{\mathbf{R}}=\mathbf{S}\mathbf{S}^{T}+\boldsymbol{\Sigma}$?
An answer to this question is important because, as is well known,
the minimizer of the function in (\ref{eq:14}) wrt a general (unrestricted)
$\mathbf{R}$ is $\mathbf{R}=\widehat{\mathbf{R}}$ (assuming that
$\widehat{\mathbf{R}}\succ0$). Therefore, if $\widehat{\mathbf{R}}$
can be written in the form (\ref{eq:3}) then the estimate of $\mathbf{R}$
obtained by minimizing (\ref{eq:14}) wrt $\mathbf{S}\mathbf{S}^{T}$
and $\boldsymbol{\Sigma}$ is $\widehat{\mathbf{R}}$ itself (of course,
the same is true for the Frobenius norm loss in (\ref{eq:5})). From
the previous analysis of $n_{m}$ and $n_{c}$, we can intuitively
expect that for $r>r_{L}$ it may be possible to write $\widehat{\mathbf{R}}$
in the form (\ref{eq:3}) (and not possible if $r<r_{L}$). The results
in \cite{shapiro} confirm this intuition, but with the following
caveat: $\widehat{\mathbf{R}}$ can indeed be almost always (i.e.
with probability one) be written in the form (\ref{eq:3}) for an
$r>r_{L}$, but the said $r$ can be significantly larger than $r_{L}$
and it is usually unknown.

A consequence of the previous discussion is that considering an $r>r_{L}$
makes little sense if the goal is to obtain a better estimate of the
covariance matrix than $\widehat{\mathbf{R}}$: indeed, for $r>r_{L}$
we might end-up with $\widehat{\mathbf{S}\mathbf{S}^{T}}+\widehat{\boldsymbol{\Sigma}}=\widehat{\mathbf{R}}$
and therefore we are back to square one. If on the other hand, the
goal is to additively decompose $\widehat{\mathbf{R}}$ as a low-rank
matrix, $\mathbf{S}\mathbf{S}^{T}$, and a diagonal one, $\boldsymbol{\Sigma}$,
then considering a rank $r>r_{L}$ might seem to make sense, but we
should keep in mind that for such an $r$ the decomposition is almost
surely not unique and therefore its use for any practical purpose
may be questioned.

The discussion in the previous paragraph has some relevance for Frisch
problem \cite{frisch} \cite{Tryphon}: find the minimum $r$ for
which $\widehat{\mathbf{R}}=\mathbf{S}\mathbf{S}^{T}+\boldsymbol{\Sigma}$
($\mathbf{S}\mathbf{S}^{T}\succeq0$ and $\mathbf{\Sigma}\succeq0$).
It follows from what was said above that generically we must have
$r>r_{L}$ and for such an $r$ the decomposition is not unique. This
simple observation suggests that the Frisch problem should be reformulated
as an approximation of $\widehat{\mathbf{R}}$ rather than an exact
decomposition: find the smallest $r$ for which $\widehat{\mathbf{R}}\approx\mathbf{S}\mathbf{S}^{T}+\boldsymbol{\Sigma}$
and the approximation error is acceptably small in a sense that is
important to the application in hand. A solution to this modified
Frisch problem can be obtained using FAAN and BIC (the Bayesian Information
Criterion), see Sections \ref{sec:4} and \ref{sec:5}. For a different
approach and algorithm, see \cite{ciccone}.

At the end of this section, we will present another lower bound on
the rank $r$ that makes it possible to write $\widehat{\mathbf{R}}$
in the form (\ref{eq:3}). From equation (\ref{eq:69}) in Appendix
B we have that 
\[
\textrm{diag}\left(\widehat{\mathbf{R}}^{-1}\right)\preccurlyeq\boldsymbol{\Sigma}^{-1}
\]
or equivalently 
\begin{equation}
\boldsymbol{\Sigma}\preccurlyeq\left[\textrm{diag}\left(\widehat{\mathbf{R}}^{-1}\right)\right]^{-1}\label{eq:E}
\end{equation}
Under the assumption that $\widehat{\mathbf{R}}$ can be written as
in (\ref{eq:3}) for a certain rank $r$, we have that 
\begin{equation}
\mathbf{S}\mathbf{S}^{T}=\left[\widehat{\mathbf{R}}-\left[\textrm{diag}\left(\widehat{\mathbf{R}}^{-1}\right)\right]^{-1}\right]+\left[\left[\textrm{diag}\left(\widehat{\mathbf{R}}^{-1}\right)\right]^{-1}-\boldsymbol{\Sigma}\right]\label{eq:52}
\end{equation}
where the second matrix on the right-hand side (rhs) is positive semi-definite
(see \eqref{eq:E}). Let $\lambda_{k1}\geq\lambda_{k2}\geq\cdots\geq\lambda_{kn}$
(for $k=1,2,3$) denote the eigenvalues of the three matrices in (\ref{eq:52})
starting with the left one. Using Weyl inequality, we deduce that
\begin{equation}
\lambda_{1i}\geq\lambda_{2i}+\lambda_{3n}\geq\lambda_{2i}\;\;\left(\textrm{for}\;i=1,\ldots,n\right)\label{eq:53-1}
\end{equation}
where the second inequality follows from the fact that $\left\{ \lambda_{3n}\geq0\right\} $.
Because $\lambda_{1i}=0$ for $i=r+1,\ldots,n$ we have from (\ref{eq:53-1})
that $\lambda_{2i}\leq0$ for $i=r+1,\ldots,n$ and therefore that
the matrix $\widehat{\mathbf{R}}-\left[\textrm{diag}\left(\widehat{\mathbf{R}}^{-1}\right)\right]^{-1}$
has at most $r$ strictly positive eigenvalues: 
\begin{equation}
r\geq n_{+}\left(\widehat{\mathbf{R}}-\left[\textrm{diag}\left(\widehat{\mathbf{R}}^{-1}\right)\right]^{-1}\right)\triangleq r_{{G}}\label{eq:gut}
\end{equation}
Above $n_{+}\left(\cdot\right)$ denotes the number of positive eigenvalues
of the argument, and the index $G$ indicates the fact that it was
the author of \cite{guttman} who introduced the bound. Note that
$r_{{G}}$ is a data-dependent lower bound and thus it applies to
every realization unlike $r_{{L}}$ that is only generically valid.

We compare $r_{G}$ to $r_{L}$ in the next example.

\emph{Example 3. Comparison of the lower bounds on the rank}

To study the tightness of a lower bound on $r$ we need to generate
matrices $\hat{\mathbf{R}}$ that can be written in the form \eqref{eq:3}
for a known rank. An important result on the Frisch decomposition
states that if $(\hat{\mathbf{R}}^{-1})_{ij}\geq0\text{ }(i,j=1,...,n)$
then the minimum rank for which $\hat{\mathbf{R}}$ can be written
in the form \eqref{eq:3} is $r=n-1$ (see, e.g., \cite{Tryphon}
and references therein). We randomly generate matrices $\hat{\mathbf{R}}$
(for $n\in[5,50]$) that satisfy the above condition, and compute
$\lceil r_{G}\rceil$ using \eqref{eq:gut} ($\lceil\cdot\rceil$
denotes the round-up operator). Fig.3 compares the average $\lceil r_{G}\rceil$
(computed from $100$ Monte-Carlo runs) with $\lceil r_{L}\rceil$
and the true rank $r=n-1$. As one can see the data-dependent lower
bound $r_{G}$ is tighter than $r_{L}$ for all values of $n$.

\begin{figure}[ht]
\begin{centering}
\begin{tabular}{c}
\includegraphics[width=8cm,height=6cm]{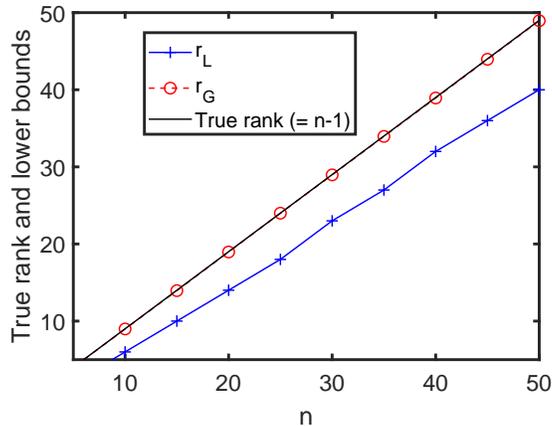} \tabularnewline
\end{tabular}
\par\end{centering}
\caption{$r_{L}$, $r_{G}$ and the true rank vs $n$. }
\end{figure}

\section{\label{sec:4}The proposed method (FAAN)}

We will use a coordinate descent algorithm to minimize the loss in
(\ref{eq:14}), which is repeated here for convenience: 
\begin{equation}
f=\textrm{Tr}\left(\hat{\mathbf{R}}\mathbf{R}^{-1}\right)+\ln\left|\mathbf{R}\right|\label{eq:22}
\end{equation}
While (\ref{eq:22}) is proportional to the negative log-likelihood
function only if the distribution of the data $\left\{ \mathbf{y}_{t}\right\} $
is normal, the above loss is a good covariance fitting criterion for
many other distributions. Note that in a misspecified case, such as
when the normal distribution assumption does not hold, (\ref{eq:22})
is usually called a quasi likelihood function (see, e.g., \cite{bai}).
\\
 Let the diagonal matrix $\boldsymbol{\Lambda}=\begin{bmatrix}\lambda_{1} &  & 0\\
 & \ddots\\
0 &  & \lambda_{r}
\end{bmatrix}\in\mathbb{R}^{r\times r}$ and the (semi) orthogonal matrix $\mathbf{U}\in\mathbb{R}^{n\times r}$
be so as 
\begin{equation}
\mathbf{U}\mathbf{\Lambda}\mathbf{U}^{T}=\boldsymbol{\Sigma}^{-1/2}(\mathbf{S}\mathbf{S}^{T})\boldsymbol{\Sigma}^{-1/2}\label{eq:23}
\end{equation}
Hence $\mathbf{U}\mathbf{\Lambda}\mathbf{U}^{T}$ is the eigenvalue
decomposition (EVD) of the rhs of (\ref{eq:23}). We reparametrize
the loss in (\ref{eq:22}) using the variables ($\mathbf{U},\boldsymbol{\Lambda}\text{ and }\boldsymbol{\Sigma}$):
\begin{equation}
f=\textrm{Tr}\left(\tilde{\mathbf{R}}\left(\boldsymbol{\Sigma}\right)\left(\mathbf{I}+\mathbf{U}\mathbf{\Lambda}\mathbf{U}^{T}\right)^{-1}\right)+\ln\left|\mathbf{I}+\mathbf{U}\mathbf{\Lambda}\mathbf{U}^{T}\right|+\ln|\boldsymbol{\Sigma}|\label{eq:24}
\end{equation}
where $\tilde{\mathbf{R}}$ is the following function of $\boldsymbol{\Sigma}$,
\begin{equation}
\tilde{\mathbf{R}}\left(\boldsymbol{\Sigma}\right)=\boldsymbol{\Sigma}^{-1/2}\hat{\mathbf{R}}\boldsymbol{\Sigma}^{-1/2}
\end{equation}
To simplify the notation, in what follows, we will use $\tilde{\mathbf{R}}$
instead of $\tilde{\mathbf{R}}\left(\boldsymbol{\Sigma}\right)$.
As we will show shortly using the reparametrization via ($\mathbf{U},\boldsymbol{\Lambda},\boldsymbol{\Sigma}$),
instead of the more traditional parametrization via $\mathbf{S}\mathbf{S}^{T}$
and $\boldsymbol{\Sigma}$, simplifies the minimization of (\ref{eq:24})
w.r.t. $\boldsymbol{\Sigma}$, for given $\mathbf{U}\textrm{ and }\boldsymbol{\Lambda}$
(see step 2 below). If we reverted from $\mathbf{U}\textrm{ and }\boldsymbol{\Lambda}$
to $\mathbf{S}\mathbf{S}^{T}$ (see (\ref{eq:23})), the minimization
of (\ref{eq:24}) w.r.t. would be more complicated (see, e.g., \cite{joreskog}). 
\begin{enumerate}
\item First we minimize (\ref{eq:24}) wrt $\mathbf{U}\mathbf{\Lambda}\mathbf{U}^{T}$,
for given $\boldsymbol{\Sigma}$. To do so, let $\mathbf{V}\in\mathbb{R}^{n\times(n-r)}$
be such that the matrix $\begin{bmatrix}\mathbf{U} & \mathbf{V}\end{bmatrix}$is
orthogonal, and observe that : 
\begin{equation}
\mathbf{I}+\mathbf{U}\mathbf{\Lambda}\mathbf{U}^{T}=\begin{bmatrix}\mathbf{U} & \mathbf{V}\end{bmatrix}\begin{bmatrix}\mathbf{I}+\mathbf{\Lambda} & \boldsymbol{0}\\
\boldsymbol{0} & \mathbf{I}
\end{bmatrix}\begin{bmatrix}\mathbf{U}^{T}\\
\mathbf{V}^{T}
\end{bmatrix}\triangleq\mathbf{Q}\mathbf{D}\mathbf{Q}^{T}
\end{equation}
Using the above EVD in (\ref{eq:24}) we can rewrite the expression
for $f$ as follows: 
\begin{equation}
f=\textrm{Tr}\left[\left(\mathbf{Q}^{T}\tilde{\mathbf{R}}\mathbf{Q}\right)\mathbf{D}^{-1}\right]+\ln|\mathbf{D}|+\ln|\boldsymbol{\Sigma}|\label{eq:27}
\end{equation}
Let $\mu_{1}\geq\mu_{2}\geq\cdots\geq\mu_{n}$ denote the eigenvalues
of $\tilde{\mathbf{R}}$, which are also the eigenvalues of $\mathbf{Q}^{T}\tilde{\mathbf{R}}\mathbf{Q}$.
Using Ruhe lower bound on the trace of the product of two symmetric
matrices (see \cite{ruhe} and also Appendix A) yields the following
inequality: 
\begin{equation}
\textrm{Tr}\left[\left(\mathbf{Q}^{T}\tilde{\mathbf{R}}\mathbf{Q}\right)\mathbf{D}^{-1}\right]\geq\sum_{k=1}^{r}\frac{\mu_{k}}{1+\lambda_{k}}+\sum_{k=r+1}^{n}\mu_{k}\label{eq:28}
\end{equation}
where the equality holds if 
\begin{equation}
\mathbf{Q}^{T}\tilde{\mathbf{R}}\mathbf{Q}=\begin{bmatrix}\mu_{1} & 0 & \cdots & 0\\
0 & \mu_{2} &  & \vdots\\
\vdots &  & \ddots & 0\\
0 & \cdots & 0 & \mu_{n}
\end{bmatrix}
\end{equation}
In particular this implies that the minimizer $\mathbf{U}$ of (\ref{eq:27})
is given by: 
\begin{equation}
\hat{\mathbf{U}}=\textrm{the matrix whose columns are the }r\textrm{ principal eigenvectors of }\tilde{\mathbf{R}}\label{eq:30}
\end{equation}
Furthermore, it follows from (\ref{eq:28}) that the minimizers \{$\lambda_{k}\geq0\}$
of (\ref{eq:27}) are the solutions to the following problems (for
$k=1,\ldots,r$): 
\begin{equation}
\begin{array}{ll}
\min\limits _{\lambda_{k}\geq0}\left[\frac{\mu_{k}}{1+\lambda_{k}}+\ln(1+\lambda_{k})\right]\end{array}\label{eq:31}
\end{equation}
The function in (\ref{eq:31}), let us say $h(\lambda_{k})$, has
a unique (unconstrained) minimum at $\lambda_{k}=\mu_{k}-1$. Indeed,
a simple calculation shows that: 
\[
h'(\lambda_{k})=-\frac{\mu_{k}}{(1+\lambda_{k})^{2}}+\frac{1}{1+\lambda_{k}}=\frac{\lambda_{k}-\left(\mu_{k}-1\right)}{\left(1+\lambda_{k}\right)^{2}}
\]
and 
\[
h''(\lambda_{k})\Bigr|_{\lambda_{k}=\mu_{k}-1}=\frac{1}{\mu_{k}^{2}}>0
\]
Therefore, the minimizer of (\ref{eq:31}) is equal to $\mu_{k}-1$
if $\mu_{k}\geq1$. If $\mu_{k}<1$, then the infimum of (\ref{eq:31})
is attained at $\lambda_{k}=0$, which follows from the fact that
in this case $h'(\lambda_{k})>0$ for $\lambda_{k}>0$. Combining
these two observations shows that the minimizer of (\ref{eq:31})
is given by ($k=1,\ldots,r$): 
\begin{equation}
\hat{\lambda}_{k}=\begin{cases}
\mu_{k}-1 & \text{if}\ \mu_{k}\geq1\\
0 & \text{else}
\end{cases}\label{eq:32}
\end{equation}
\item In the second step of FAAN we minimize (\ref{eq:24}) wrt $\boldsymbol{\Sigma}$,
for given $\mathbf{U}$ and $\boldsymbol{\Lambda}$.

Let 
\begin{equation}
\boldsymbol{\Gamma}=(\mathbf{I}+\mathbf{U}\mathbf{\Lambda}\mathbf{U}^{T})^{-1}
\end{equation}
We need to solve the problem 
\begin{equation}
\begin{array}{ll}
\min\limits _{\boldsymbol{\Sigma}}\textrm{Tr}\left(\boldsymbol{\Sigma}^{-1/2}\hat{\mathbf{R}}\boldsymbol{\Sigma}^{-1/2}\boldsymbol{\Gamma}\right)+\ln\left|\boldsymbol{\Sigma}\right|\end{array}\label{eq:func}
\end{equation}
where $\boldsymbol{\Sigma}$ is defined in (\ref{eq:2}). A straightforward
calculation shows that 
\begin{equation}
\begin{split}\frac{\partial}{\partial\sigma_{k}}\textrm{Tr}\left(\hat{\mathbf{R}}\boldsymbol{\Sigma}^{-1/2}\boldsymbol{\Gamma}\boldsymbol{\Sigma}^{-1/2}\right) & =\frac{\partial}{\partial\sigma_{k}}\sum_{i=1}^{n}\sum_{j=1}^{n}\frac{\hat{R}_{ij}\Gamma_{ij}}{\sigma_{i}\sigma_{j}}\\
 & =-2\sum_{i=1,i\neq k}^{n}\frac{\hat{R}_{ik}\Gamma_{ik}}{\sigma_{i}\sigma_{k}^{2}}-\frac{2\hat{R}_{kk}\Gamma_{kk}}{\sigma_{k}^{3}}
\end{split}
\label{eq:35}
\end{equation}
and 
\begin{equation}
\frac{\partial}{\partial\sigma_{k}}\ln|\boldsymbol{\Sigma}|=\frac{2}{\sigma_{k}}\label{eq:36}
\end{equation}
It follows from (\ref{eq:35}) and (\ref{eq:36}) that the minimizer
$\sigma_{k}$ of (\ref{eq:func}), for fixed \{$\sigma_{i}\}_{i\neq k}$,
is the positive solution to the equation: 
\begin{equation}
\sigma_{k}^{2}-\left(\sum_{i=1,i\neq k}^{n}\frac{\hat{R}_{ik}\Gamma_{ik}}{\sigma_{i}}\right)\sigma_{k}-\hat{R}_{kk}\Gamma_{kk}=0\label{root}
\end{equation}
which is given by 
\begin{equation}
\hat{\sigma}_{k}=\frac{b_{k}+\sqrt{b_{k}^{2}+4c_{k}}}{2}\label{eq:38}
\end{equation}
where 
\begin{equation}
\begin{cases}
b_{k} & =\underset{i=1,i\neq k}{\overset{n}{\sum}}\frac{\hat{R}_{ik}\Gamma_{ik}}{\sigma_{i}}\\
c_{k} & =\hat{R}_{kk}\Gamma_{kk}
\end{cases}\label{eq:39}
\end{equation}

\end{enumerate}
To verify that (\ref{eq:38}) is indeed a minimizer, note from (\ref{eq:35})
and (\ref{eq:36}) that the second-order derivative of (\ref{eq:func})
with respect to $\sigma_{k}$ is given by: 
\begin{equation}
2\left(\frac{3c_{k}}{{\bf \sigma_{k}^{4}}}+\frac{2b_{k}}{{\bf \sigma_{k}^{3}}}-\frac{1}{{\bf \sigma_{k}^{2}}}\right)\sim\;\;\;-{\bf \sigma_{k}^{2}}+2b_{k}{\bf \sigma_{k}}+3c_{k}\label{eq:N1}
\end{equation}
(where the symbol $\sim$ denotes the fact that the two expressions
have the same sign). The value of (\ref{eq:N1}) at $\sigma_{k}=\hat{\sigma}_{k}$
satisfies: 
\begin{equation}
\begin{split} & \underbrace{-\hat{\sigma}_{k}^{2}+b_{k}\hat{\sigma}_{k}+c_{k}}_{0}+b_{k}\hat{\sigma}_{k}+2c_{k}\sim b_{k}^{2}+b_{k}\sqrt{b_{k}^{2}+4c_{k}}+4c_{k}\\
 & \sim\sqrt{b_{k}^{2}+4c_{k}}+b_{k}\geq0
\end{split}
\end{equation}
and this proves that (\ref{eq:38}) is a minimum point. 
\begin{rem}
By the same analysis as above one can verify that the negative root
of (\ref{root}) is also a minimum point but of the unconstrained
minimization problem. If the negative root of (\ref{root}) leads
to a smaller value of the loss in (\ref{eq:22}) than that corresponding
to the positive root, then one may expect that the unconstrained minimum
of the negative likelihood function lies outside the feasibility set
(in which $\boldsymbol{\Sigma}\succeq0$). Interestingly, even in
such a case, FAAN (which picks up the positive root of (\ref{root}))
never leaves the feasibility set despite the fact that, unlike other
algorithms, it does not require any correction measures by the user.
Note that in such cases FAAN may converge to a point near the boundary
of the feasibility set thus approaching what is called a Heywood solution. 
\end{rem}
A summary of the FAAN algorithm follows. 
\begin{algorithm}[H]
\caption{FAAN algorithm}

\textbf{Input:} $\hat{\mathbf{R}}$, $r$ and $\hat{\boldsymbol{\Sigma}}_{o}$
(for instance $\hat{\boldsymbol{\Sigma}}_{o}=\mathbf{I}$ or $\textrm{diag}(\hat{\mathbf{R}})$
or randomly generated).

For $i=0,1,2,\ldots,$ do: 
\begin{itemize}
\item Compute the EVD of $\hat{\boldsymbol{\Sigma}}_{i}^{-1/2}\hat{\mathbf{R}}\hat{\boldsymbol{\Sigma}}_{i}^{-1/2}$
and $(\hat{\mathbf{U}}_{i},\hat{\boldsymbol{\Lambda}}_{i})$ using
(\ref{eq:30}) and (\ref{eq:32}). 
\item Compute $\hat{\boldsymbol{\Sigma}}_{i+1}$ using (\ref{eq:38}) (for
$k=1,\dots,n$) where we use the updated estimates of $\sigma_{i}$
for $i<k$ and $\widehat{\boldsymbol{\Gamma}}_{i}=(\mathbf{I}+\hat{\mathbf{U}}_{i}\hat{\mathbf{\boldsymbol{\Lambda}}_{i}}\mathbf{\hat{U}}_{i}^{T})^{-1}$.
This step can be iterated a few times or until a convergence criterion
is satisfied (note that the computational cost of iterating this step
is basically negligible). 
\item If $\left(f_{i-1}-f_{i}\right)/f_{i}\leq\epsilon$ (where $f_{-1}=\infty$
and $\epsilon=10^{-3}$ for example) then goto Output; else $i\rightarrow i+1$
and continue the iterations. 
\end{itemize}
\textbf{Output}: $\hat{\mathbf{U}},\,\,\hat{\mathbf{\boldsymbol{\Lambda}}},\,\,\hat{\boldsymbol{\Sigma}}$
and $\widehat{\mathbf{S}\mathbf{S}^{T}}=\hat{\boldsymbol{\Sigma}}^{1/2}\hat{\mathbf{U}}\hat{\mathbf{\boldsymbol{\Lambda}}}\hat{\mathbf{U}}^{T}\hat{\boldsymbol{\Sigma}}^{1/2}$. 
\end{algorithm}

\begin{figure}
\begin{centering}
\begin{tabular}{cc}
\includegraphics[width=8cm,height=6cm]{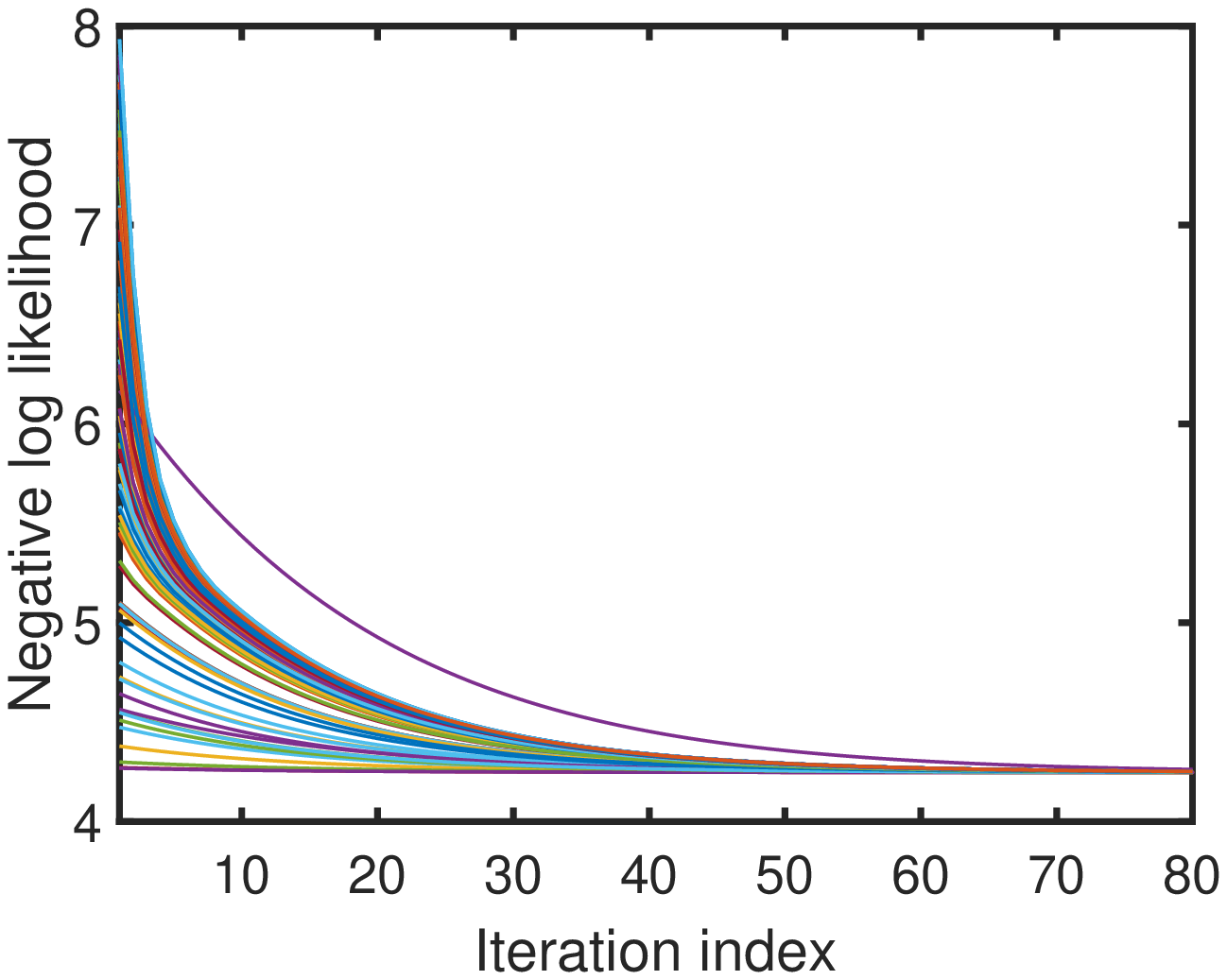} 
  & \includegraphics[width=8cm,height=6cm]{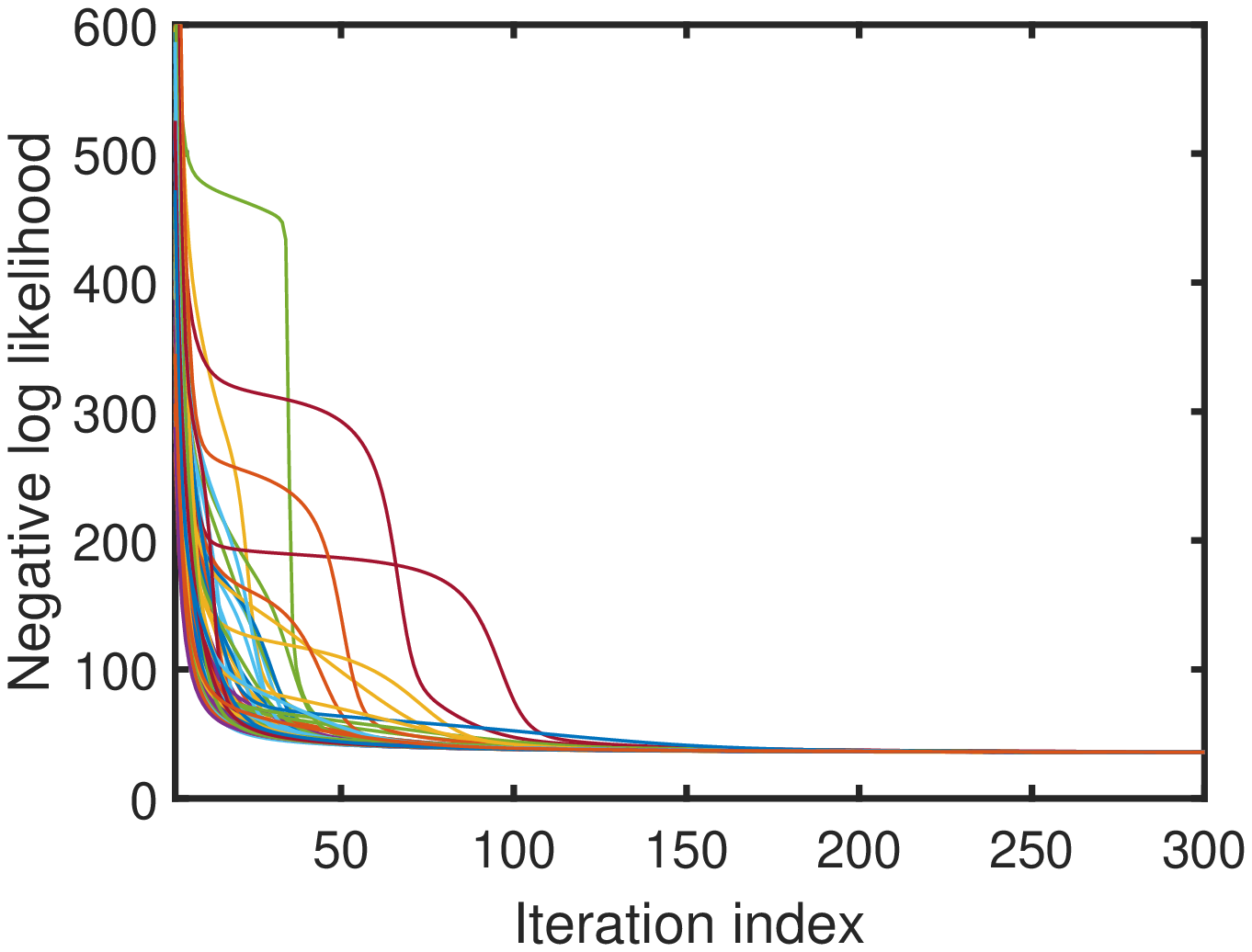}\tabularnewline
\end{tabular}
\par\end{centering}
\caption{\label{fig:3-1} Monotonic convergence behavior of FAAN for $100$
different initializations ($\hat{\mathbf{R}}$ was randomly generated).
(a) $n=10$, $N=20$, $r=4$ and (b) $n=1000$, $N=1500$, $r=100$.}
\end{figure}

\begin{rem}
The above derivation of FAAN also provides the minimizers of the likelihood
function (\ref{eq:24}) in the special case of isotropic noise: $\boldsymbol{\Sigma}=\sigma^{2}\mathbf{I}$.
In this case, $\widetilde{\mathbf{R}}=\widehat{\mathbf{R}}/\sigma^{2}$
and (\ref{eq:30}) becomes: 
\begin{equation}
\hat{\mathbf{U}}=\textrm{the matrix made from the }r\textrm{ principal eigenvectors of }\widehat{\mathbf{R}}
\end{equation}
Let $\rho_{1}\geq\rho_{2}\geq\cdots\geq\rho_{n}$ denote the eigenvalues
of $\widehat{\mathbf{R}}$. The unique minimizer of the function of
$\lambda_{k}$ in (\ref{eq:31}) still is $\hat{\lambda}_{k}=\mu_{k}-1$.
Because $\mu_{k}=\rho_{k}/\sigma^{2}$, we have 
\begin{equation}
\hat{\lambda}_{k}=\frac{\rho_{k}}{\sigma^{2}}-1\quad(k=1,\ldots,r)\label{eq:56-1}
\end{equation}
where $\sigma^{2}$ is yet to be determined. Under the assumption
that $\rho_{k}/\sigma^{2}\geq1$ at the minimum, which will be shown
to be true below, (\ref{eq:56-1}) is the minimizer of (\ref{eq:31})
also when the constraint $\lambda_{k}\geq0$ is enforced. Inserting
(\ref{eq:56-1}) in (\ref{eq:27}) and \eqref{eq:28} yields the following
function of $\sigma^{2}$: 
\begin{equation}
\begin{aligned}\frac{1}{\sigma^{2}}\underset{k=r+1}{\overset{n}{\sum}}\rho_{k}+\ln\left|\mathbf{D}\right|+\ln\left|\boldsymbol{\Sigma}\right| & =\frac{1}{\sigma^{2}}\underset{k=r+1}{\overset{n}{\sum}}\rho_{k}-r\ln\sigma^{2}+\underset{k=1}{\overset{r}{\sum}}\ln\rho_{k}+n\ln\sigma^{2}\\
 & =\frac{1}{\sigma^{2}}\underset{k=r+1}{\overset{n}{\sum}}\rho_{k}+(n-r)\ln\sigma^{2}+\textrm{const}.
\end{aligned}
\label{eq:57}
\end{equation}
The first-order derivative of (\ref{eq:57}) wrt $\sigma^{2}$ is
given by: 
\begin{equation}
\frac{n-r}{\sigma^{2}}-\frac{1}{\sigma^{4}}\underset{k=r+1}{\overset{n}{\sum}}\rho_{k}
\end{equation}
and thus the unique minimizer of (\ref{eq:57}) is: 
\begin{equation}
\widehat{\sigma}^{2}=\frac{1}{n-r}\underset{k=r+1}{\overset{n}{\sum}}\rho_{k}
\end{equation}
which obviously satisfies $\rho_{k}\geq\widehat{\sigma}^{2}$ (for
$k=1,...,r$) as required for (\ref{eq:56-1}) to hold. This observation
completes our derivation of the ML estimates in the case of $\boldsymbol{\Sigma}=\sigma^{2}\mathbf{I}$,
which appears to be simpler than the usual proof of this result in
the literature. 
\end{rem}
The convergence of FAAN to a local minimum of the loss function in
(\ref{eq:22}) follows from theoretical results on coordinate descent
(also called cyclic or alternating minimization) algorithms, see e.g.,
\cite{tseng} \cite{beck}. Our empirical experience is in agreement
with the theory: in all numerical experiments we have performed to
date, we did not find a single case in which FAAN had convergence
problems. An illustration of the monotonic convergence of FAAN is
presented in Fig. \ref{fig:3-1} for multiple random initializations
of $\boldsymbol{\hat{\Sigma}}$. 

In sum, it is our experience that the loss function in (\ref{eq:22}),
despite being non-convex, appears to have a favourable landscape which
makes its minimization by a coordinate descent method like FAAN a
relatively straightforward task.

At the end of this section, we remind the reader that if $r$ is ``sufficiently
large'' (in any case $r>\max(r_{L},r_{G}))$ then it follows from
the discussion in Section III that the estimated covariance matrix
provided by FAAN, let us say $\hat{\hat{\mathbf{R}}}$, matches $\mathbf{\hat{R}}$
exactly: 
\begin{equation}
\hat{\mathbf{R}}=\hat{\boldsymbol{\Sigma}}+\widehat{\mathbf{S}\mathbf{S}^{T}}\triangleq\hat{\hat{\mathbf{R}}}
\end{equation}
Interestingly, for \textit{any} $r$ the diagonals of $\hat{\hat{\mathbf{R}}}$
and $\hat{\mathbf{R}}$ are identical: 
\begin{equation}
\textrm{diag}(\hat{\hat{\mathbf{R}}})=\textrm{diag}(\hat{\mathbf{R}})\label{eq:55}
\end{equation}
(see Appendix B for the proof). A possible use of this result is as
a partial test for convergence.

Finally we note that the present covariance modeling exercise may
require not only estimation of $\boldsymbol{\Sigma}$ and $\mathbf{S}\mathbf{S}^{T}$
but also estimation of $r$ (if it is unknown). When estimation of
the integer parameter $r$ is required, the fact that FAAN is a ML
method is a clear advantage as we can directly make use of information
criterion rules such as BIC (see, e.g., \cite{yngve}, \cite{FDR})
to estimate $r$ along with $\boldsymbol{\Sigma}$ and $\mathbf{S}\mathbf{S}^{T}$.
A more detailed discussion of this aspect can be found in the next
section.

\section{\label{sec:5}Numerical Applications}

We will present two applications: 1) DOA estimation from the spatial
samples collected by an array of sensors, where we need to estimate
$\mathcal{R}\left(\mathbf{S}\right)$; and 2) Portfolio selection
for financial asset management, for which we aim to get a covariance
matrix estimate that is more accurate than $\hat{\mathbf{R}}.$

\subsection{DOA estimation}

We consider a uniform linear array comprising $n$ sensors on which
impinge the signals emitted by $m$ sources. The output of the array
(after a preprocessing step that includes quadrature sampling) can
be described by the FA model in (\ref{eq:1}) where now $\left\{ \mathbf{y}_{t}\right\} $,
$\left\{ \mathbf{s}_{t}\right\} $ and $\left\{ \mathbf{e}_{t}\right\} $
are complex-valued, and the matrix $\mathbf{A}$ is given by (see,
e.g., \cite{stoicamusic}): 
\begin{equation}
\footnotesize\begin{bmatrix}1 & \cdots & 1\\
e^{i2\pi f_{1}} & \cdots & e^{i2\pi f_{m}}\\
\vdots & \ddots & \vdots\\
e^{i2\pi f_{1}(n-1)} & \cdots & e^{i2\pi f_{m}(n-1)}
\end{bmatrix}\label{eq:42}
\end{equation}
The DOAs of the sources can be readily obtained from the spatial frequencies
$\left\{ f_{k}\right\} $ in (\ref{eq:42}) (see, e.g., the cited
paper), therefore the problem is to estimate $\left\{ f_{k}\right\} $.

\begin{figure}[ht]
\begin{centering}
\begin{tabular}{cc}
\includegraphics[width=8cm,height=6cm]{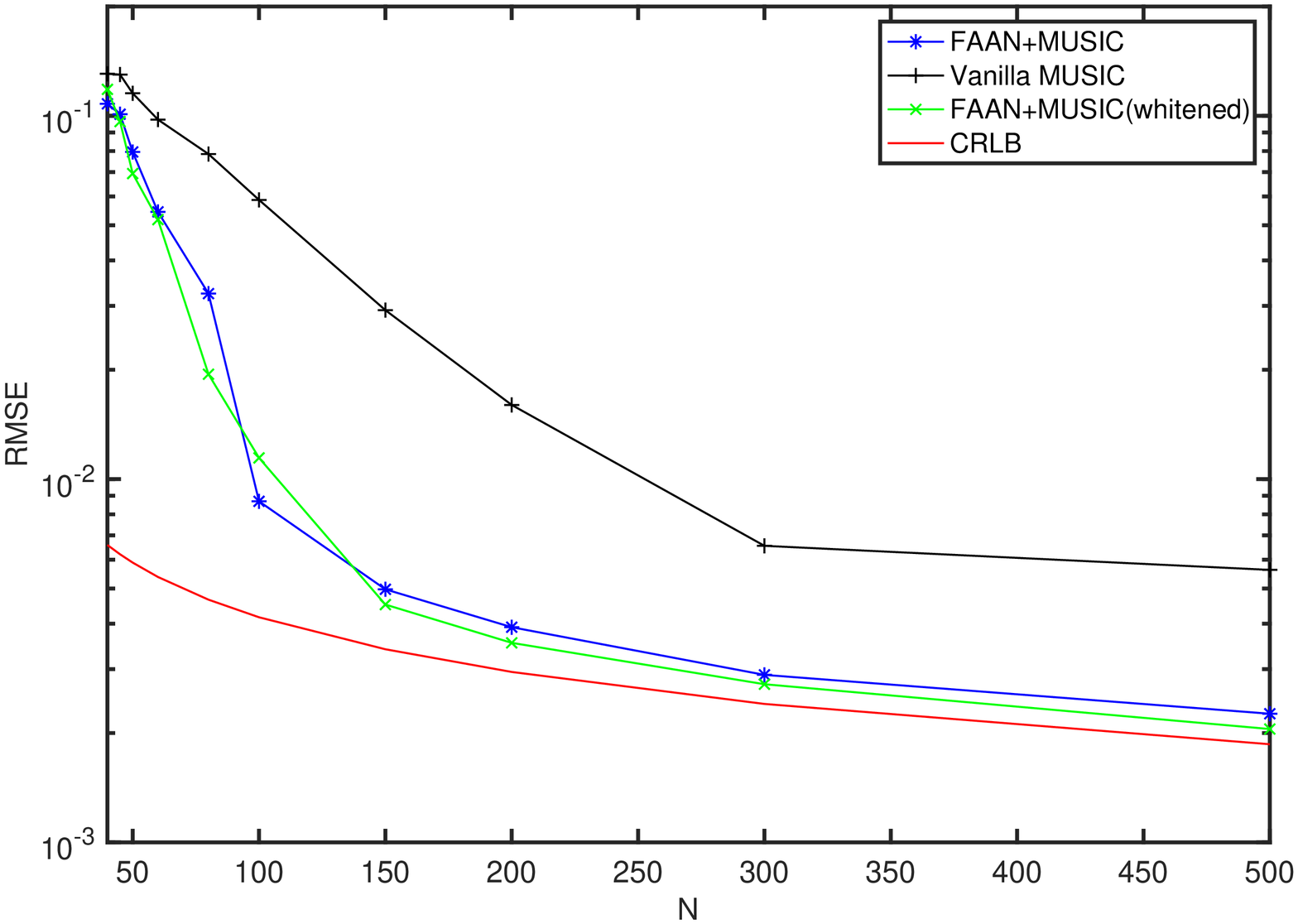}  & \includegraphics[width=8cm,height=6cm]{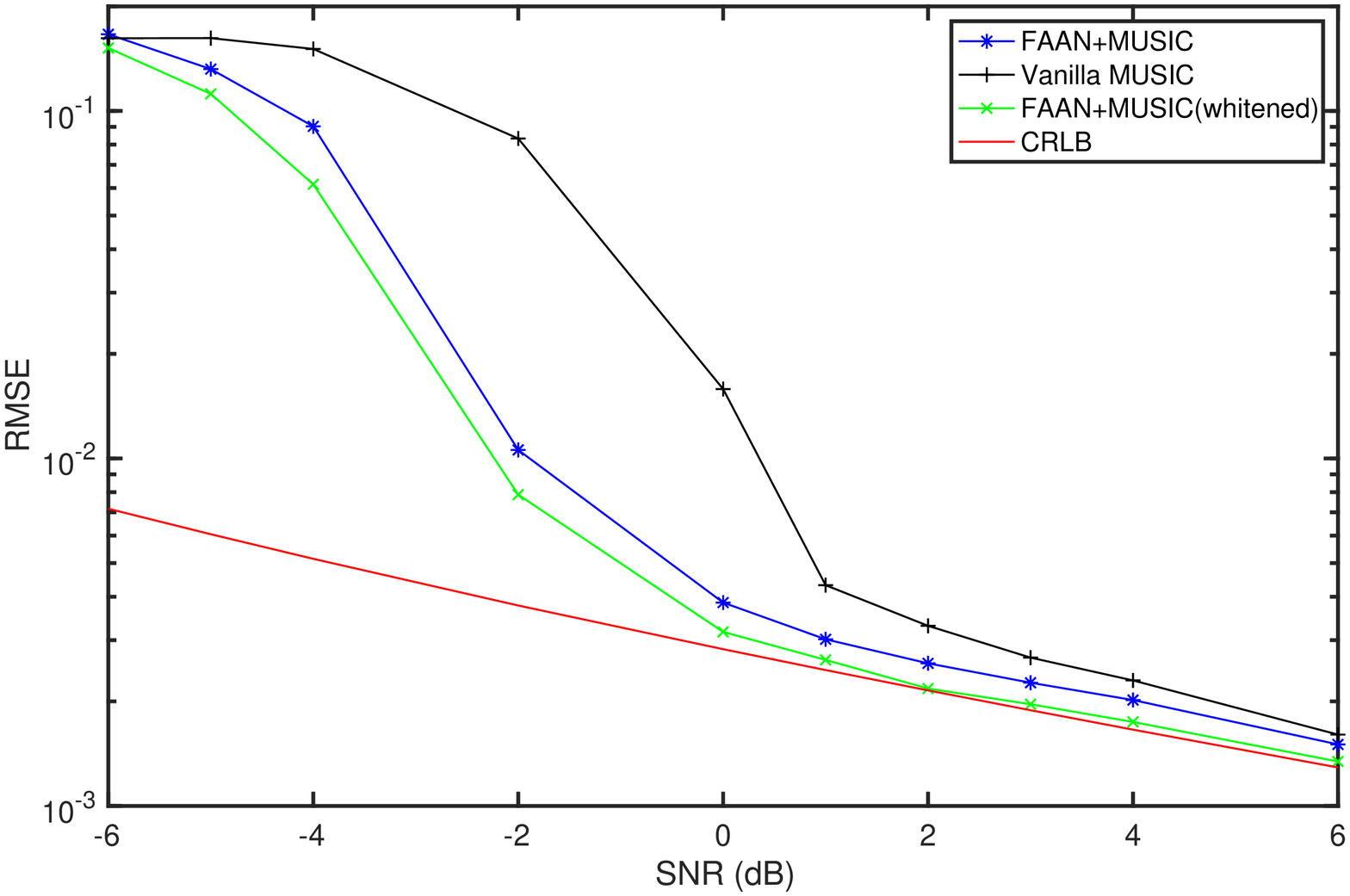}\tabularnewline
a) RMSE vs $N$ (SNR= $0$ dB)  & b) RMSE vs SNR ($N=80$)\tabularnewline
\end{tabular}
\par\end{centering}
\caption{RMSE curves for the DOA estimation application.}
\label{fig:3} 
\end{figure}

The FAAN algorithm presented in the previous section can be directly
applied to the complex-valued data with the only modification being
that $b_{k}$ in (\ref{eq:39}) should be defined as the real part
of the rhs. However, to keep the discussion within the framework of
real-valued FA considered in the previous sections, we will use only
the real part of $\left\{ \mathbf{y}_{t}\right\} $, for which all
variables in the FA model (\ref{eq:1}) are real-valued and the $\mathbf{A}$
matrix is defined as: 
\begin{equation}
\footnotesize\mathbf{A}=\begin{bmatrix}\cos\left(2\pi f_{1}0\right) & \sin\left(2\pi f_{1}0\right) & \cdots & \cos\left(2\pi f_{m}0\right) & \sin\left(2\pi f_{m}0\right)\\
\vdots & \vdots & \vdots & \vdots & \vdots\\
\cos\left(2\pi f_{1}(n-1)\right) & \sin\left(2\pi f_{1}(n-1)\right) & \cdots & \cos\left(2\pi f_{m}(n-1)\right) & \sin\left(2\pi f_{m}(n-1)\right)
\end{bmatrix}\label{eq:43}
\end{equation}
We generate data with \eqref{eq:1} and (\ref{eq:43}) where 
\begin{itemize}
\item $n=15$, $m=2$, $f_{1}=0.2$, $f_{2}=0.25$. 
\item The elements of the signals $\left\{ \mathbf{s}_{t}\right\} $ are
independently drawn from a standard normal distribution. 
\item The $k$th element of the noise vector $\left\{ \mathbf{e}_{t}\right\} $
is drawn from a normal distribution with mean zero and variance $\sigma_{k}^{2}$;
$\left\{ \sigma_{k}^{2}\right\} $ are independently drawn from a
uniform distribution on $(0,1)$ and then fixed and also scaled so
as the signal to noise ratio (SNR), 
\[
\textrm{SNR}\triangleq10\log_{10}\left(\frac{\textrm{Tr}\left(\mathbf{A}\mathbf{A}^{T}\right)}{\sum_{k=1}^{n}\sigma_{k}^{2}}\right)\:\text{dB}
\]
takes a desired value. 
\item $N$ and SNR are varied in the intervals: 
\begin{align*}
N & \in\left[40,500\right]\\
\textrm{SNR} & \in\left[-6,6\right]\textrm{dB}
\end{align*}
\end{itemize}
We run FAAN for each dataset generated as described above and use
$\hat{\mathbf{S}}=\hat{\boldsymbol{\Sigma}}^{1/2}\hat{\mathbf{U}}\hat{\boldsymbol{\Lambda}}^{1/2}$
(with $r=4$) as an estimate of a basis of $\mathcal{R}(\mathbf{S})$.
Then we compute the estimates of $\{f_{k}\}$ using the MUSIC (MUltiple
SIgnal Classification) algorithm (see, e.g., \cite{liao}, \cite{stoicamusic}
and references therein) applied to the FAAN estimate of $\mathbf{R}$
and also to the whitened $\hat{\mathbf{R}}$, i.e. $\hat{\boldsymbol{\Sigma}}^{-1/2}\hat{\mathbf{R}}\hat{\boldsymbol{\Sigma}}^{-1/2}$,
where $\hat{\boldsymbol{\Sigma}}$ denotes the estimate of ${\boldsymbol{\Sigma}}$
provided by FAAN. In the former case, the MUSIC estimates of $\{f_{k}\}_{k=1}^{2}$
are given by the locations of the two largest peaks of the following
function: 
\[
\footnotesize\left\Vert \left(\hat{\mathbf{S}}^{T}\hat{\mathbf{S}}\right)^{-1/2}\hat{\mathbf{S}}^{T}\begin{bmatrix}\cos\left(2\pi f0\right) & \sin\left(2\pi f0\right)\\
\vdots & \vdots\\
\cos\left(2\pi f(n-1)\right) & \sin\left(2\pi f(n-1)\right)
\end{bmatrix}\right\Vert 
\]
In the latter case, the MUSIC estimates of $\{f_{k}\}$ are obtained
as follows. Let $\hat{\mathbf{U}}_{\textrm{S}}$ denote the estimate
of the signal subspace given by the four principal eigenvectors of
$\hat{\boldsymbol{\Sigma}}^{-1/2}\hat{\mathbf{R}}\hat{\boldsymbol{\Sigma}}^{-1/2}$.
The estimates of $\{f_{k}\}$ are then obtained as the locations of
the two largest peaks of the function: 
\[
\footnotesize\left\Vert \hat{\mathbf{U}}_{\text{S}}^{T}\hat{\boldsymbol{\Sigma}}^{-1/2}\begin{bmatrix}\cos\left(2\pi f0\right) & \sin\left(2\pi f0\right)\\
\vdots & \vdots\\
\cos\left(2\pi f(n-1)\right) & \sin\left(2\pi f(n-1)\right)
\end{bmatrix}\right\Vert 
\]
Using $500$ independent noise realizations, we estimate the average
root mean square error (RMSE) of the frequency estimates 
\[
\textrm{RMSE}=\frac{1}{2}\sum_{k=1}^{2}\sqrt{\mathbb{E}(\hat{f}_{k}-f_{k})^{2}}
\]
Figures \ref{fig:3}(a) and \ref{fig:3}(b) show the variation of
RMSE versus $N$ and versus SNR, respectively. We have also included
the Cram{�}r-Rao Lower Bounds (CRLB) in these figures, which have
been computed using the Slepian-Bangs formula (see Appendix B in \cite{SASbook}).
As one can see, the RMSE of the MUSIC estimates obtained using FAAN
approach the CRLB as $N$ or SNR increases and they are significantly
lower than those of the vanilla MUSIC algorithm which estimates the
signal subspace directly from the sample covariance matrix.

To conclude this application, we make use of the bound $r_{L}$ to
determine the number of sources that can be uniquely resolved with
an array of $n$ sensors. In the case of isotropic noise and for the
matrix $\mathbf{A}$ in (\ref{eq:43}) we have 
\begin{equation}
m\leq\frac{n}{2}-0.5\label{eq:44}
\end{equation}
For anisotropic noise, it follows from the discussion in Section III
that $m$ must generically satisfy : 
\begin{equation}
m=\frac{r}{2}\leq r_{L}\slash2=\frac{n}{2}+0.25\left(1-\sqrt{8n+1}\right)\label{eq:45}
\end{equation}
Fig. 6 compares the integer parts of rhs in (\ref{eq:44}) and (\ref{eq:45}).
As expected, the maximum number of sources that can be resolved from
data corrupted by anisotropic noise is smaller than in the case of
isotropic noise, but the relative difference between these two cases
decreases as $n$ increases.

\begin{figure}
\begin{centering}
\includegraphics[width=8cm,height=6cm]{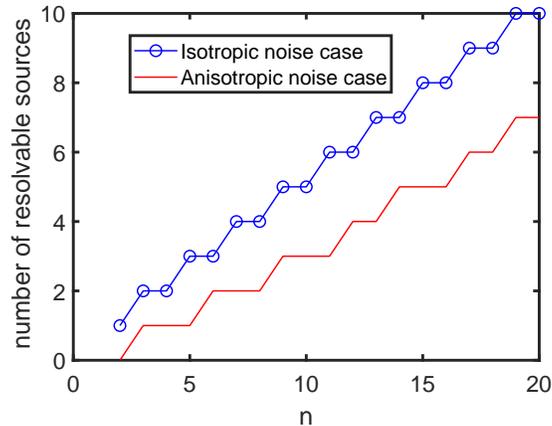} 
\par\end{centering}
\label{fig:5} \caption{The number of resolvable sources vs $n$ for the isotropic and anisotropic
noise cases.}
\end{figure}

\subsection{Portfolio selection}

In this subsection we will discuss a portfolio selection application
in which the assumption is that the covariance matrix of the stock
returns satisfies the model in (\ref{eq:3}). This assumption is not
unrealistic as the number of independent market factors that are internally
driving the stock prices (and also their returns) is usually quite
small. Consequently the data equation in (\ref{eq:1}) is well suited
to model stock returns, the covariance matrix of which will therefore
follow the FA model in (\ref{eq:3}). However, compared to DOA estimation,
the portfolio selection problem is more challenging as the number
of samples available to estimate $\mathbf{R}$ is usually small and
in some cases $\hat{\mathbf{R}}$ is even rank-deficient. In such
cases in which $N$ is small, the MLE (ML estimate) of $\mathbf{R}$
may not exist. The paper \cite{rob} has derived the necessary and
sufficient conditions for the existence of MLE: 
\begin{itemize}
\item[a)] For $N\geq n$ the MLE of $\mathbf{R}$ (see \eqref{eq:3} and \eqref{eq:14})
exists with probability one. 
\item[b)] For $N<n$ but $N\geq r$ the MLE also exists with probability one. 
\item[c)] For $N<r$, the negative log-likelihood function in (\ref{eq:14})
is unbounded below, so the MLE does not exist. 
\end{itemize}
In the portfolio selection problem the number of data samples is larger
than that of the factors, thus usually we are in the safe cases a)
or b).

Before applying FAAN to financial data, we study its performance on
synthetic data that mimic the real-world data. We simulate normal
data samples of dimension $n=40$ whose covariance matrix (which was
generated randomly) satisfies the model in (\ref{eq:3}) with $r=3$
and SNR $=0$ dB. We run FAAN on the simulated data for different
values of $N$. In practical applications in addition to the estimation
of $\mathbf{S}\mathbf{S}^{T}$ and $\boldsymbol{\Sigma}$ we also
need to estimate the number of factors $r$. We will estimate $r$
using BIC: 
\begin{equation}
\hat{r}=\mathop{\arg\min}\limits _{r\in[1,10]}\text{BIC}(r)
\end{equation}
where 
\begin{equation}
\text{BIC}(r)=N\left(\textrm{Tr}(\hat{\mathbf{R}}(\widehat{\mathbf{S}\mathbf{S}^{T}}+\hat{\boldsymbol{\Sigma}})^{-1})+\ln{|\widehat{\mathbf{S}\mathbf{S}^{T}}+\hat{\boldsymbol{\Sigma}}|}\right)+n_{m}\ln{(Nn)}
\end{equation}
and where $n_{m}$ is the number of real-valued model parameters,
see \eqref{eq:16}. In Fig. 7a we plot Prob($\hat{r}=r$) versus $N$.
As can be seen from this figure BIC can find the true rank with a
probability close to one even when the number of samples $N$ is less
than $n$. In Fig. 7b we show the normalized $\textrm{RMSE}\triangleq\frac{\|\mathbf{R}_{\textrm{true}}-\hat{\mathbf{R}}_{\textrm{alg}}\|}{\|\mathbf{R}_{\textrm{true}}\|}$
where $\hat{\mathbf{R}}_{\textrm{alg}}$ denotes either the SCM or
the FAAN estimate of ${\mathbf{R}}$. It can be seen that FAAN outperforms
the SCM by about $10\%$. 
\begin{figure}
\begin{centering}
\begin{tabular}{cc}
\label{fig:5} \includegraphics[width=8cm,height=6cm]{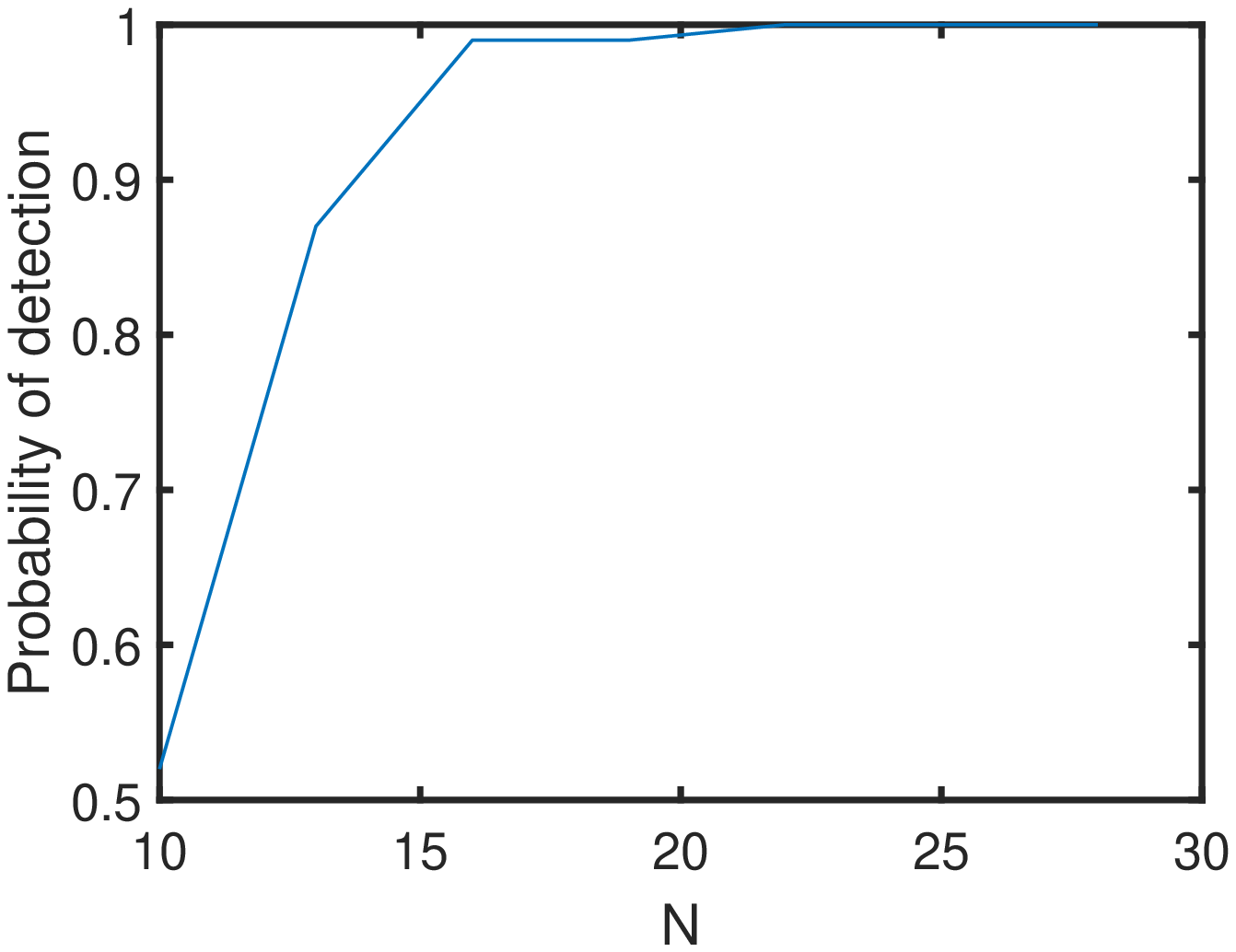}  & \includegraphics[width=8cm,height=6cm]{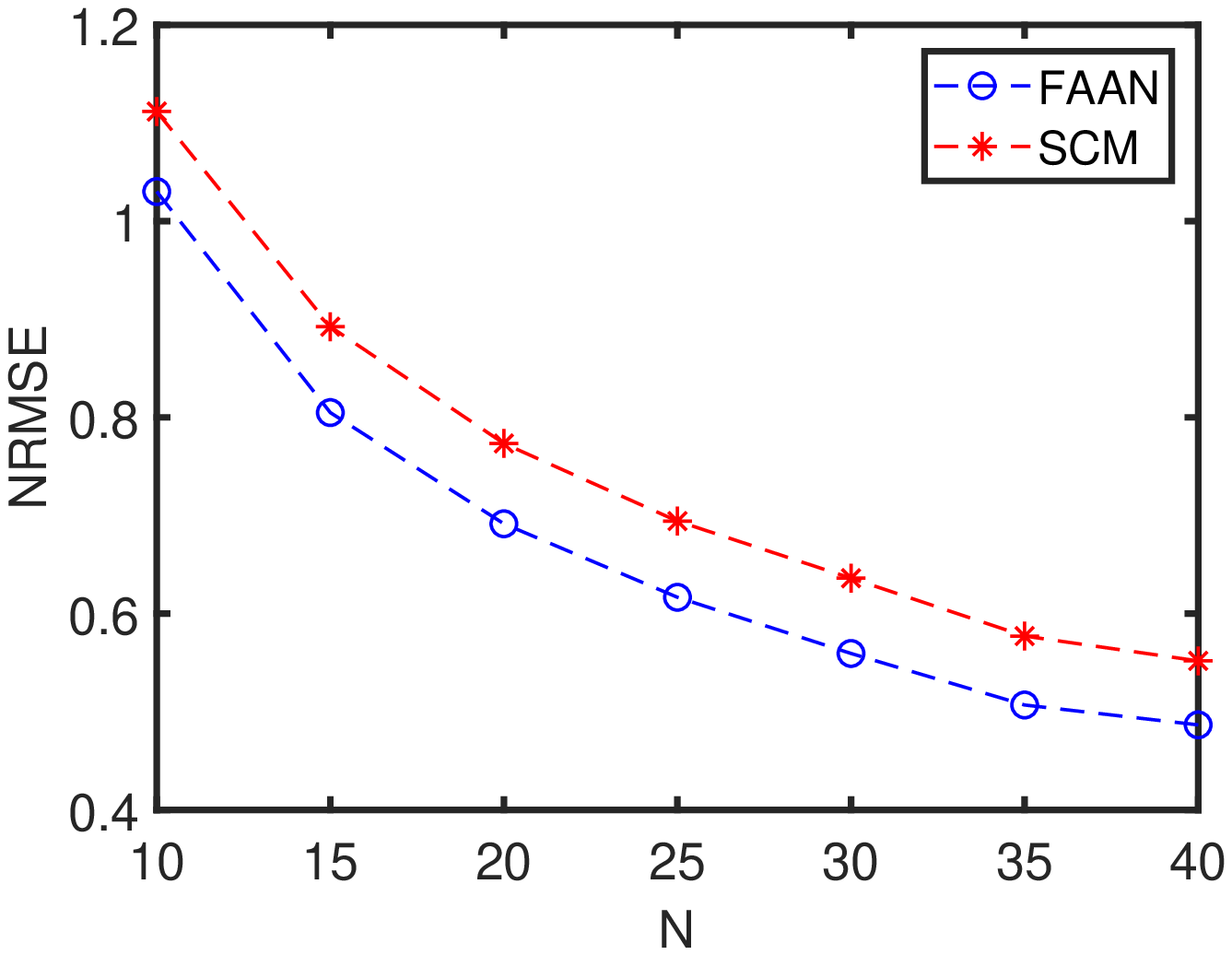}\tabularnewline
(a) Probability of detection vs $N$  & (b) NRMSE vs $N$ \tabularnewline
\end{tabular}
\par\end{centering}
\caption{Synthetic example mimicking financial data ($n=40$ and the true rank
is $r=3$).}
\end{figure}

\begin{figure}[ht]
\begin{centering}
\begin{tabular}{c}
\includegraphics[width=8cm,height=6cm]{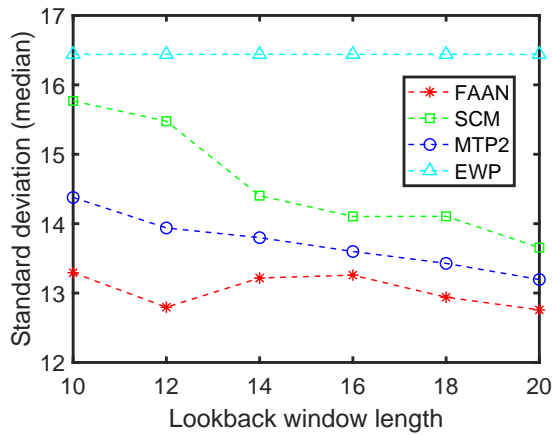}\tabularnewline
\end{tabular}
\par\end{centering}
\caption{\label{fig:Financial-data}Financial data analysis.}
\end{figure}

Next, we present the real-world financial data analysis. We will use
the daily stock returns of $n=40$ assets over a period of $7580$
days from the publicly available database of the Center for Research
in Security Prices \cite{CRSP}. The samples $\left\{ \mathbf{y}_{t}\right\} $
are obtained from raw stock market data as in \cite{MTP2}. The stock
market data corresponds to $360$ investment dates at intervals of
one month, where each month comprises $20$ trading days. Using the
most recent $N$ returns, we first estimate the covariance matrix
$\mathbf{R}_{d}$ corresponding to each investment date ($d=1,2,\ldots,360$)
using FAAN and BIC ($r$ is also unknown and needs to be estimated
in this application). We then solve the following minimum variance
portfolio problem : 
\begin{equation}
\begin{alignedat}{1}\underset{\mathbf{w}_{d}\in\mathbb{R}^{n}}{\min} & \mathbf{w}_{d}^{T}\mathbf{R}_{d}\mathbf{w}_{d},\\
\textrm{s.t. } & \mathbf{w}_{d}^{T}\mathbf{1}=1
\end{alignedat}
\label{eq:25}
\end{equation}
to obtain the portfolio weight vector $\mathbf{w}_{d}$ for the $d$-th
month. The standard derivation of the portfolio returns is the performance
metric since a good estimator of the covariance matrix will give a
small value of the portfolio variance ($\mathbf{w}_{d}^{T}\mathbf{R}_{d}\mathbf{w}_{d}$).
The standard deviation corresponding to each month is computed using
the next $4$ months ($84$ days) out-of-sample (future) returns.
We use the equally weighted portfolio (EWP) as a baseline for comparison,
and consider the SCM and a recently proposed method called MTP2 \cite{MTP2}
as competing estimators. The lookback window length $N$ is varied
between $10$ and $20$, and the median of the standard deviations
corresponding to different months is computed. Fig. \ref{fig:Financial-data}
shows the median vs the lookback window length for each method. FAAN
outperforms the other methods especially in the case of small $N$.

\section{Conclusions}

We have presented a new method for Factor Analysis in the case of
Anisotropic Noise (FAAN). The proposed method is a coordinate descent
algorithm that monotonically maximizes the normal likelihood function,
is easy to understand as well as implement (in particular it requires
no tuning by the user), and has excellent convergence properties.
Compared to FAAN, most of the existing algorithms are either more
complex (both computationally and conceptually) and in need of some
tuning by the user or are less reliable (for instance, they can converge
to points outside the feasibility set, or may not converge at all).
The following aspects, which were deemed to be important for FAAN
and its use in practical applications, have been discussed: i) identifiability
of the covariance matrix model (i.e., uniqueness of parametrization);
ii) exact matching of the sample covariance matrix as the number of
factor increases; iii) data independent/dependent lower bounds on
the number of factors that are important for the previous two aspects;
and iv) estimation of the number of factors using the Bayesian Information
Criterion (BIC).

Out of the many possible applications of FAAN, we have selected two:
a) DOA estimation from the signals collected by an array of sensors;
and b) portfolio selection for financial asset management. As shown
in the paper, FAAN has performed well in both cases. In the era of
``big (high-dimensional) data'', applications that need dimensionality
reduction techniques are abundant and the hope is that FAAN will be
found to be a useful tool for FA in many other practical cases besides
those considered in this paper. A MATLAB code for FAAN can be downloaded
from \url{https://github.com/ghaniafatima/FAAN}

\appendix

\subsection{Ruhe trace lower bound}

To make the paper as self-contained as possible, we state in this
appendix a generalized version of Ruhe lower bound \cite{ruhe} and
also provide a simple proof of it.

Let $\mathbf{A}\in\mathbb{R}^{n\times n}$ and $\mathbf{B}\in\mathbb{R}^{n\times n}$
be two symmetric matrices with eigenvalues $\lambda_{1}\left(\mathbf{A}\right)\geq\lambda_{2}\left(\mathbf{A}\right)\geq\ldots\geq\lambda_{n}\left(\mathbf{A}\right)$
and $\lambda_{1}\left(\mathbf{B}\right)\geq\lambda_{2}\left(\mathbf{B}\right)\geq\ldots\geq\lambda_{n}\left(\mathbf{B}\right)$
respectively. Then, 
\begin{equation}
\textrm{Tr}\left(\mathbf{A}\mathbf{B}\right)\geq\sum_{i=1}^{n}\lambda_{i}\left(\mathbf{A}\right)\lambda_{n+1-i}\left(\mathbf{B}\right)\label{eq:46}
\end{equation}
Evidently the above inequality holds (with equality) for $n=1$. To
prove the inequality in the general case we show that if it holds
for $n-1$ then it also holds for $n$. Let 
\[
\mathbf{A}=\mathbf{E}\boldsymbol{\Lambda}_{\mathbf{A}}\mathbf{E}^{T}\;\;\;\;\;\;\;\;\boldsymbol{\Lambda}_{\mathbf{A}}=\begin{bmatrix}\lambda_{1}\left(\mathbf{A}\right) & 0 & \cdots & 0\\
0 & \lambda_{2}\left(\mathbf{A}\right) &  & \vdots\\
\vdots &  & \ddots & 0\\
0 & \cdots & 0 & \lambda_{n}\left(\mathbf{A}\right)
\end{bmatrix}
\]
denotes the EVD of $\mathbf{A}$, and let $\mathbf{C}=\mathbf{E}^{T}\mathbf{B}\mathbf{E}$.
Using this notation, we can write: 
\begin{equation}
\begin{aligned}\begin{split}\textrm{Tr}\left(\mathbf{A}\mathbf{B}\right)\end{split}
 & =\textrm{Tr}\left(\boldsymbol{\Lambda}_{\mathbf{A}}\mathbf{C}\right)=\sum_{i=1}^{n-1}\lambda_{i}\left(\mathbf{A}\right)C_{ii}+\lambda_{n}\left(\mathbf{A}\right)C_{nn}\\
 & =\sum_{i=1}^{n-1}\lambda_{i}\left(\mathbf{A}\right)C_{ii}+\lambda_{n}(\mathbf{A})\left[\textrm{Tr}(\mathbf{C})-\sum_{i=1}^{n-1}C_{ii}\right]
\end{aligned}
\end{equation}
\begin{align}
 & =\sum_{i=1}^{n-1}\left(\lambda_{i}\left(\mathbf{A}\right)-\lambda_{n}\left(\mathbf{A}\right)\right)C_{ii}+\lambda_{n}\left(\mathbf{A}\right)\textrm{Tr}(\mathbf{C})\label{eq:48}
\end{align}
The first term in $(\ref{eq:48})$ is the trace of the product of
a diagonal matrix with elements $\left\{ \lambda_{i}\left(\mathbf{A}\right)-\lambda_{n}\left(\mathbf{A}\right)\right\} $
and an $(n-1)\times(n-1)$ matrix denoted $\bar{\mathbf{C}}$ that
is a principal block of $\mathbf{C}=\begin{bmatrix}\bar{\mathbf{C}} & \cdot\\
. & .
\end{bmatrix}$.

By Cauchy interlace theorem, we have that: 
\[
\lambda_{i}\left(\mathbf{C}\right)\geq\lambda_{i}\left(\bar{\mathbf{C}}\right)\geq\lambda_{i+1}\left(\mathbf{C}\right),\;\;\;\;i=1,\ldots,n-1
\]
Using this fact together with the assumption that (\ref{eq:46}) holds
for $(n-1)\times(n-1)$ matrices we deduce that: 
\begin{equation}
\begin{split}\textrm{Tr}\left(\mathbf{A}\mathbf{B}\right)\geq & \sum_{i=1}^{n-1}\left(\lambda_{i}\left(\mathbf{A}\right)-\lambda_{n}\left(\mathbf{A}\right)\right)\lambda_{n-i}\left(\bar{\mathbf{C}}\right)+\lambda_{n}\left(\mathbf{A}\right)\textrm{Tr}(\mathbf{C})\\
 & \geq\sum_{i=1}^{n-1}\left(\lambda_{i}\left(\mathbf{A}\right)-\lambda_{n}\left(\mathbf{A}\right)\right)\lambda_{n+1-i}\left(\mathbf{C}\right)+\lambda_{n}(\mathbf{A})\sum_{i=1}^{n}\lambda_{n+1-i}(\mathbf{C})\\
 & =\sum_{i=1}^{n}\lambda_{i}\left(\mathbf{A}\right)\lambda_{n+1-i}\left(\mathbf{B}\right)
\end{split}
\end{equation}
(in last equality above we used the fact that $\mathbf{C}$ and $\mathbf{B}$
have the same eigenvalues). The lower-bound in (\ref{eq:46}) is thus
proved.

\subsection{Proof of the diagonal matching property}

The stationary points of the loss function (\ref{eq:22}) are solutions
of the equations: 
\begin{equation}
\frac{\partial f}{\partial\Sigma_{ii}}=0,\;\;\;\;\;\frac{\partial f}{\partial S_{ij}}=0\;\;\;\;i,j=1,\ldots,n
\end{equation}
To calculate the above derivatives, we use the fact that for a matrix
$\mathbf{B}(x)$ which is a function of a variable $x\in\mathbb{R}$
, the following formulas hold true: 
\begin{equation}
\begin{aligned} & \frac{\partial}{\partial x}\left[\ln\mathbf{B}(x)\right]=\textrm{Tr}\left[\mathbf{B}^{-1}(x)\frac{\partial\mathbf{B}(x)}{\partial x}\right]\\[3pt]
 & \frac{\partial}{\partial x}\left[\mathbf{B}^{-1}(x)\right]=-\mathbf{B}^{-1}(x)\frac{\partial\mathbf{B}(x)}{\partial x}\mathbf{B}^{-1}(x)
\end{aligned}
\end{equation}
Making use of these formulas we get after some straightforward calculations
(below $\boldsymbol{\alpha}_{i}$ denotes the $i^{\textrm{th}}$ column
of the identity matrix):\\

\begin{equation}
\frac{\partial f}{\partial\Sigma_{ii}}=-\textrm{Tr}\left(\hat{\mathbf{R}}\mathbf{R}^{-1}\boldsymbol{\alpha}_{i}\boldsymbol{\alpha}_{i}^{T}\mathbf{R}^{-1}\right)+\textrm{Tr}\left(\mathbf{R}^{-1}\boldsymbol{\alpha}_{i}\boldsymbol{\alpha}_{i}^{T}\right)=0
\end{equation}
or equivalently, 
\begin{equation}
\textrm{diag}\left[\mathbf{R}^{-1}\left(\mathbf{R}-\hat{\mathbf{R}}\right)\mathbf{R}^{-1}\right]=0\label{eq:53}
\end{equation}
and 
\[
\frac{1}{2}\frac{\partial f}{\partial S_{ij}}=-\textrm{Tr}\left(\hat{\mathbf{R}}\mathbf{R}^{-1}\mathbf{S}\boldsymbol{\alpha}_{i}\boldsymbol{\alpha}_{j}^{T}\mathbf{R}^{-1}\right)+\textrm{Tr}\left(\mathbf{R}^{-1}\mathbf{S}\boldsymbol{\alpha}_{j}\boldsymbol{\alpha}_{j}^{T}\right)=0
\]
or equivalently 
\begin{equation}
\left(\mathbf{R}-\hat{\mathbf{R}}\right)\mathbf{R}^{-1}\mathbf{S}=\boldsymbol{0}\label{eq:54}
\end{equation}
By the matrix inversion lemma, 
\begin{equation}
\left(\mathbf{S}\mathbf{S}^{T}+\boldsymbol{\Sigma}\right)^{-1}=\boldsymbol{\Sigma}^{-1}-\boldsymbol{\Sigma}^{-1}\mathbf{S}(\mathbf{I}+\mathbf{S}^{T}\boldsymbol{\Sigma}^{-1}\mathbf{S})^{-1}\mathbf{S}^{T}\boldsymbol{\Sigma}^{-1}\label{eq:69}
\end{equation}
which implies that: 
\begin{equation}
\mathbf{R}^{-1}\mathbf{S}=\boldsymbol{\Sigma}^{-1}\mathbf{S}\left(\mathbf{I}+\mathbf{S}^{T}\boldsymbol{\Sigma}^{-1}\mathbf{S}\right)^{-1}\left[\left(\mathbf{I}+\mathbf{S}^{T}\boldsymbol{\Sigma}^{-1}\mathbf{S}\right)-\mathbf{S}^{T}\boldsymbol{\Sigma}^{-1}\mathbf{S}\right]=\boldsymbol{\Sigma}^{-1}\mathbf{S}\left(\mathbf{I}+\mathbf{S}^{T}\boldsymbol{\Sigma}^{-1}\mathbf{S}\right)^{-1}\label{eq:56}
\end{equation}
Using (\ref{eq:54}) in (\ref{eq:56}) yields 
\begin{equation}
\left(\mathbf{R}-\hat{\mathbf{R}}\right)\boldsymbol{\Sigma}^{-1}\mathbf{S}=0
\end{equation}
which together with (\ref{eq:53}) and \eqref{eq:69} show that,

\begin{equation}
\textrm{diag}\left[\mathbf{R}^{-1}\left(\mathbf{R}-\hat{\mathbf{R}}\right)\mathbf{R}^{-1}\right]=\textrm{diag}\left[\boldsymbol{\Sigma}^{-1}\left(\mathbf{R}-\hat{\mathbf{R}}\right)\boldsymbol{\Sigma}^{-1}\right]=0
\end{equation}
and the proof of \eqref{eq:55} is concluded.

 \bibliographystyle{ieeetr}
\bibliography{cov_est_factormodel}

\end{document}